\documentclass[%
 prl, reprint,
 superscriptaddress,
 nofootinbib,   % keep Supplemental Material footnotes at the foot of the page
 nobibnotes,
 amsmath,amssymb,
 aps, physrev,
]{revtex4-2}

\usepackage[dvipsnames]{xcolor}
\usepackage{soul}
\usepackage{verbatim}
\usepackage{comment}
\usepackage{graphicx}% Include figure files
\usepackage{epstopdf}
\usepackage{float}%    provides the [H] placement used in the Supplement
\usepackage{dcolumn}%  Align table columns on decimal point
\usepackage{array}
\usepackage{booktabs}
\usepackage{multirow}
\usepackage{amsfonts}
\usepackage{bm}%       bold math
\usepackage{bbm}
\usepackage{dsfont}
\usepackage[utf8]{inputenc}
\usepackage[normalem]{ulem}
\definecolor{linkblue}{RGB}{0,70,140}
\definecolor{citegreen}{RGB}{0,110,90}
\definecolor{urlpurple}{RGB}{120,40,140}
\usepackage[
    colorlinks=true,
    linkcolor=linkblue,
    citecolor=citegreen,
    urlcolor=urlpurple
]{hyperref}

\newcommand{\hchi}{\hat\chi}
\newcommand{\hq}{\hat q_0}
\newcommand{\be}{\begin{equation}}
\newcommand{\ee}{\end{equation}}
\newcommand{\bea}{\begin{eqnarray}}
\newcommand{\eea}{\end{eqnarray}}
\newcommand{\beas}{\begin{eqnarray*}}
\newcommand{\eeas}{\end{eqnarray*}}

\newcommand{\dd}{{\mathrm d}}

\newcommand{\re}{{\rm e}}

\begin{document}

%\preprint{APS/123-QED}

\title{\textbf{Large deviations for linear regressions} 
}% 

\author{Silvia Bartolucci}
\email{s.bartolucci@ucl.ac.uk}
\affiliation{Department of Computer Science, University College London,  WC1E 6EA, London (UK)}
\author{Fabio Caccioli}
\affiliation{Department of Computer Science, University College London,  WC1E 6EA, London (UK)}
\author{Francesco Caravelli}
\affiliation{T-Division (T-4), Los Alamos National Laboratory, Los Alamos NM, 87544 (USA)}
%\affiliation{INeP, Lucca 55012, IT}%\affiliation{INeP, Lucca 55012, (IT)}

\author{Pierpaolo Vivo}
\affiliation{Department of Mathematics, King’s College London, WC2R 2LS, London (UK)}



\begin{abstract}
Linear regression is one of the simplest and most widely used tools to learn patterns from data: it
fits a set of coefficients so that a linear combination of predictors best matches observed responses.
The quality of the fit is measured by the residual sum of squares, the total squared mismatch
between predictions and data, whose minimum defines the training loss. We consider Gaussian design and
noise, with teacher coefficients independently drawn from a general distribution $p(\beta)$, and a general class of separable regularizers,
including Ridge and Lasso. Using the zero-temperature replica method, we
compute analytically the large-deviation statistics of the minimum training loss for large numbers $P$ of
predictors and $N$ of observations, with  $r=P/N$ fixed. The rate function we compute governs rare sample-to-sample fluctuations of the optimal loss. Extensive numerical simulations are in excellent agreement with our theory and clearly show a
pronounced deviation from the Gaussian regime of typical fluctuations in the tails.
\end{abstract}


\maketitle

%\tableofcontents
\textit{Introduction - } Linear and generalized linear modelling, often implemented through
least-squares or likelihood-based fitting, are among the most fundamental tools
in data analysis, used across the natural and social sciences to extract signal from noisy
measurements. In physics, such methods underlie very common experimental
procedures, from fitting spectral lines \cite{Juvela2024SPIF,Noll1980Voigt}
and calibrating detectors \cite{Knoll2010RadiationDetection,Ribeiro2005GammaCalibration}
to inferring parameters in cosmological surveys
\cite{Planck2020CosmologicalParameters,DESI2025BAO}. Related linear-model
frameworks are also central in high-dimensional experimental data analysis,
including neuroimaging \cite{Friston1994GLM} and high-throughput genomics
\cite{Smyth2005Limma,Ritchie2015Limma,Love2014DESeq2}. More broadly, these
models appear wherever one seeks to explain an observed quantity as a weighted linear
combination of measured features.

\begin{figure}[htb!]
    \centering
    \includegraphics[width=\linewidth]{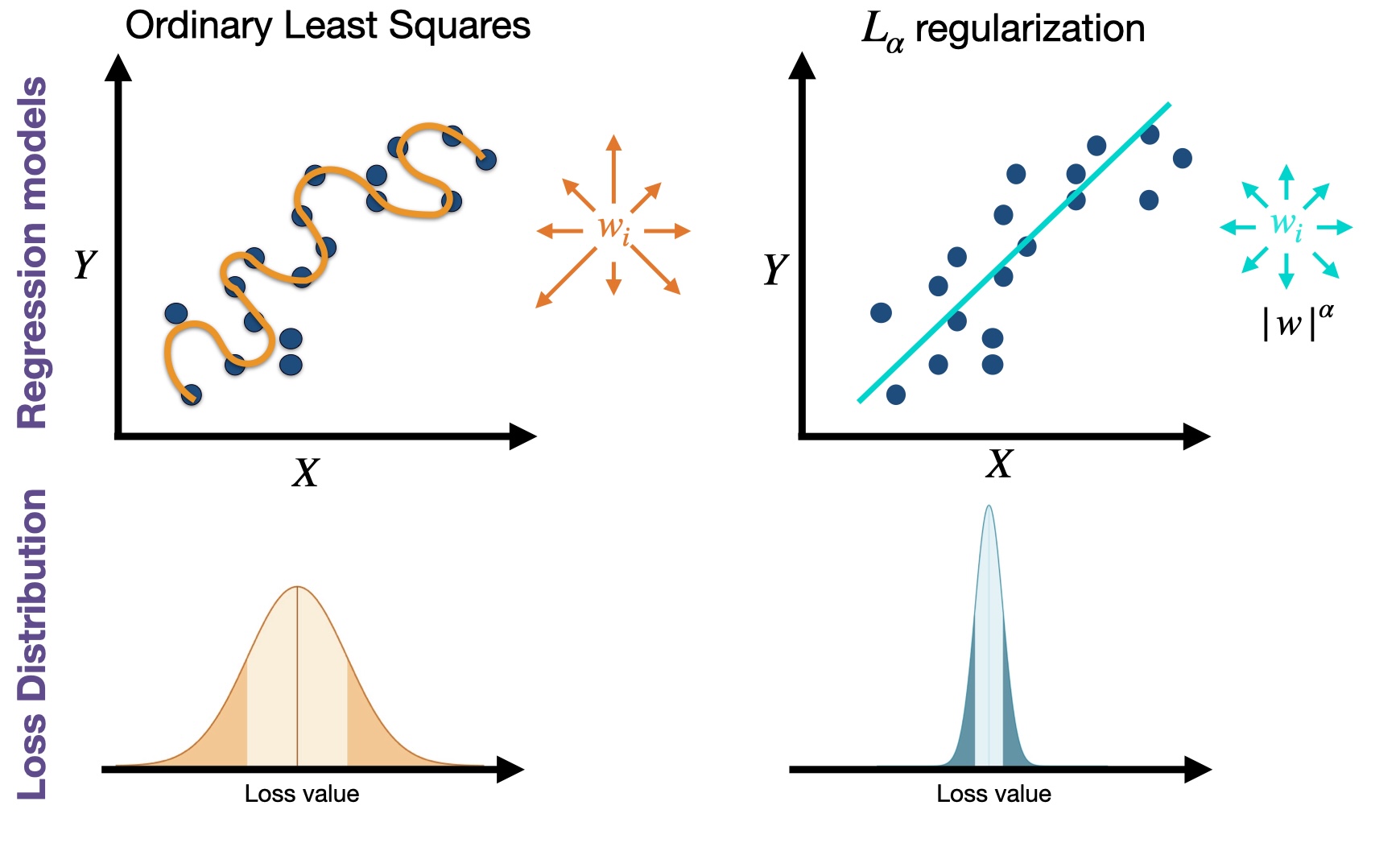}
    \caption{Schematic representation of regression models in proximity of the interpolation threshold with and without $L_\alpha$ regularization, associated weights $w_i$'s and distribution of the training loss.}
    \label{fig:scheme}
\end{figure}

Given a set of $N$ observations collected
in a response vector $\bm y \in \mathbb{R}^N$, and a matrix of $P$ explanatory
variables $X \in \mathbb{R}^{N \times P}$ --- which might represent, for example,
experimental conditions, basis functions, or molecular descriptors --- one seeks
coefficients $\bm w \in \mathbb{R}^P$ such that $X\bm w/\sqrt{P}$ approximates $\bm y$ as closely
as possible. The natural measure of fit is the \emph{residual sum of squares},
\begin{equation}
    \mathrm{RSS}(\bm w) = |\bm y - X\bm w/\sqrt{P}|^2\ ,\label{defRSSeq1}
\end{equation}
which sums the squared discrepancies between the model's predictions and the actual
observations. Because this cost is computed on the same data used to determine
$\bm w$, it is called the \emph{training loss}, and minimizing it gives the best fit to the
training data (see scheme in Fig. \ref{fig:scheme}).

Despite this elementary formulation, linear regression becomes a rich and
physically interesting problem when the number of model parameters $P$ is
comparable to (or even exceeds) the number of data points $N$. This
\emph{high-dimensional regime} arises naturally in modern science: in genomics,
genome-wide studies routinely involve hundreds of thousands to millions of
genetic variants, motivating sparse and high-dimensional regression methods for
large-scale genetic prediction problems
\cite{Tibshirani1996LASSO,Guan2011BVSR,Wray2013Pitfalls};
in compressed sensing, one reconstructs sparse or compressible signals from
far fewer measurements than ambient unknowns
\cite{Candes2006Robust,Donoho2006CompressedSensing};
and in neural-network theory, overparameterized models with more parameters
than training examples are central to modern questions about interpolation,
generalization, benign overfitting, and the double-descent phenomenon
\cite{Zhang2017Rethinking,Belkin2019DoubleDescent,
Bartlett2020BenignOverfitting,Jacot2018NTK}. 

Consider now a scientific experiment in which a limited number of samples is collected and a high-dimensional regression model is fitted using hundreds of candidate predictors. Even when the underlying signal and noise levels are fixed statistically, the particular finite sample can make the inference problem unusually easy or difficult: some datasets happen to constrain the relevant directions well, while others are nearly degenerate or poorly informative. This scenario is common. In statistics, screening many predictors can generate apparently strong fits even under weak or absent signal \cite{Freedman1983,FreedmanPee1989,FanShaoZhou2018,FanZhou2016}; in polygenic prediction, regularized models are fitted to very large numbers of genetic variants using finite cohorts \cite{Wray2013Pitfalls,Kullo2022PolygenicScores}; and in data-driven materials physics, large descriptor spaces are searched for compact models of experimentally relevant properties \cite{Ghiringhelli2015BigData,Ouyang2018SISSO,Bartel2018Gibbs}. In all these settings, the scientific conclusion is ultimately based on one finite realization of the data, so it is important to know not only the typical performance of the estimator, but also how likely unusually favorable or unfavorable fits are to occur \cite{CawleyTalbot2010Overfitting}. Typical asymptotic theory predicts the representative `cost' of fitting such a model, but characterizing rare departures from this typical behavior is a longstanding and substantially harder problem, especially beyond the textbook regime where all the sources of disorder are Gaussian. In this Letter, we overcome this challenge using an enhanced version of the classical replica method from the statistical physics of disordered systems, which allows us to precisely quantify how likely the inference problem is to be unusually easy or unusually difficult to fit.

Statistical physics approaches are well suited to analyzing linear regression in the high-dimensional regime by treating the data as random variables and characterizing the resulting fluctuations across realizations. Near the interpolation threshold---where the model first becomes capable of fitting the training data exactly---quantities such as the prediction error, the norm of the fitted coefficients, and the estimation error exhibit sharp behavior reminiscent of phase transitions in disordered systems \cite{Kondor2008Divergent,Barbier2019OptimalErrors,Hastie2022Ridgeless}. These phenomena have motivated extensive analytical work on the typical behavior of high-dimensional estimators using random-matrix theory, high-dimensional asymptotics, and statistical mechanics \cite{Kondor2008Divergent,Zdeborova2016StatPhysInference,DobribanWager2018HighDim,Hastie2022Ridgeless,AdvaniGanguli2016,Coolen2020Overfitting,Mozeika2021Exact,Mei2022RandomFeatures,BarbierPanchenkoSaenz2022,BarbierChenPanchenkoSaenz2025}.
Similar phase-transition phenomena have also been identified in closely related high-dimensional optimization problems, such as portfolio optimization under various risk measures, where statistical mechanics methods have revealed analogous estimation instabilities and the stabilizing role of regularization \cite{Kondor2008Divergent,Pafka2002Noisy,Kondor2007Noise,Ciliberti2007ExpectedShortfall,Still2010Regularizing,VargaHaszonits2016Replica}. 

In high-dimensional regression, increasing the number of parameters relative to the number of observations may lead to overfitting.
A classical remedy to overfitting is \emph{regularization}: the
addition of penalty terms that bias the solution towards simpler or more
physically motivated forms. Ridge regression adds an $\ell_2$ penalty on the
coefficient magnitudes, stabilizing directions poorly constrained by the data
and shrinking the fitted coefficients towards zero
\cite{HoerlKennard1970Ridge,Bishop2006PRML}. Lasso regression instead adds an
$\ell_1$ penalty, which promotes sparse solutions by driving many coefficients
to exactly zero \cite{Tibshirani1996LASSO}, a particularly natural choice
when one expects only a small number of the $P$ candidate variables to be
genuinely relevant \cite{ParkCasella2008BayesianLasso}. Both methods have been studied
extensively through \emph{typical} quantities such as average training error,
prediction error, and the retrieval accuracy 
\cite{DobribanWager2018HighDim,BayatiMontanari2012LASSORisk,
Wainwright2009LassoThresholds}, with fewer results on atypical behavior in Refs.~\cite{Rush2020LassoRate,ZhuLu2026,CuiSagliettiZdeborova2021}. 

For definiteness, we now introduce the precise model we work with.\\

\textit{Model - }
We consider the standard linear regression model with $N$ observations and $P$ predictors. Let $x_j^{(i)}$ denote the $j$-th feature for the $i$-th data point, and $y^{(i)}$ the corresponding response variable, with $i=1,\ldots,N$ and $j=1,\ldots,P$. The model assumes that the observed response is a linear combination of the input features, with additive Gaussian noise. 
The input features $x_j^{(i)}$ are drawn independently from the standard normal distribution $x_j^{(i)} \sim \mathcal{N}(0,1)$, 
and the response variable $y^{(i)}$ is generated according to the model
\begin{equation}
y^{(i)}[\bm{\epsilon,x}] = \frac{1}{\sqrt{P}}\sum_{j} \beta_j x_j^{(i)} + \epsilon^{(i)}\ , 
\end{equation}
with additive noise $\epsilon^{(i)}$ drawn from a standard normal distribution $\epsilon^{(i)} \sim \mathcal{N}(0,1)$. We assume that the parameters $\beta_j$ are independently drawn from a general, potentially non-Gaussian probability density function $p(\beta)$.

The optimal training loss is the combination of the residual sum of squares in Eq. \eqref{defRSSeq1} with the chosen regularizer $g_\lambda(w)$ and can be therefore written in the form
\begin{equation}
    H_\lambda(\bm w)=\frac{1}{2} 
\sum_{i=1}^{N}
\left(
y^{(i)} - \sum_{j=1}^{P} \frac{w_j}{\sqrt{P}} x_j^{(i)}
\right)^2 +\sum_{j=1}^P g_\lambda(w_j)\ .
\end{equation}
In the terminology of the statistical mechanics of learning, $\beta_j$ defines the ``teacher weights'' generating the responses, while $w_j$ denotes the ``student weights'' fitted from the sample. Most common choices for the regularizer are $g_\lambda(w)=(\lambda/2) w^2$ (Ridge) and $g_\lambda(w)=(\lambda/2) |w|$ (Lasso), while Ordinary Least Squares (OLS) corresponds to $g_\lambda\equiv 0$. The factor $1/\sqrt{P}$ is introduced so that, for weights of order $O(1)$, the term 
$\sum_{j} \frac{w_j}{\sqrt{P}} x_j^{(i)}$ remains of order $O(1)$, as $y^{(i)}$ and the cost scale linearly with $N$. 

\begin{figure}
    \centering
    \includegraphics[width=0.99\linewidth]{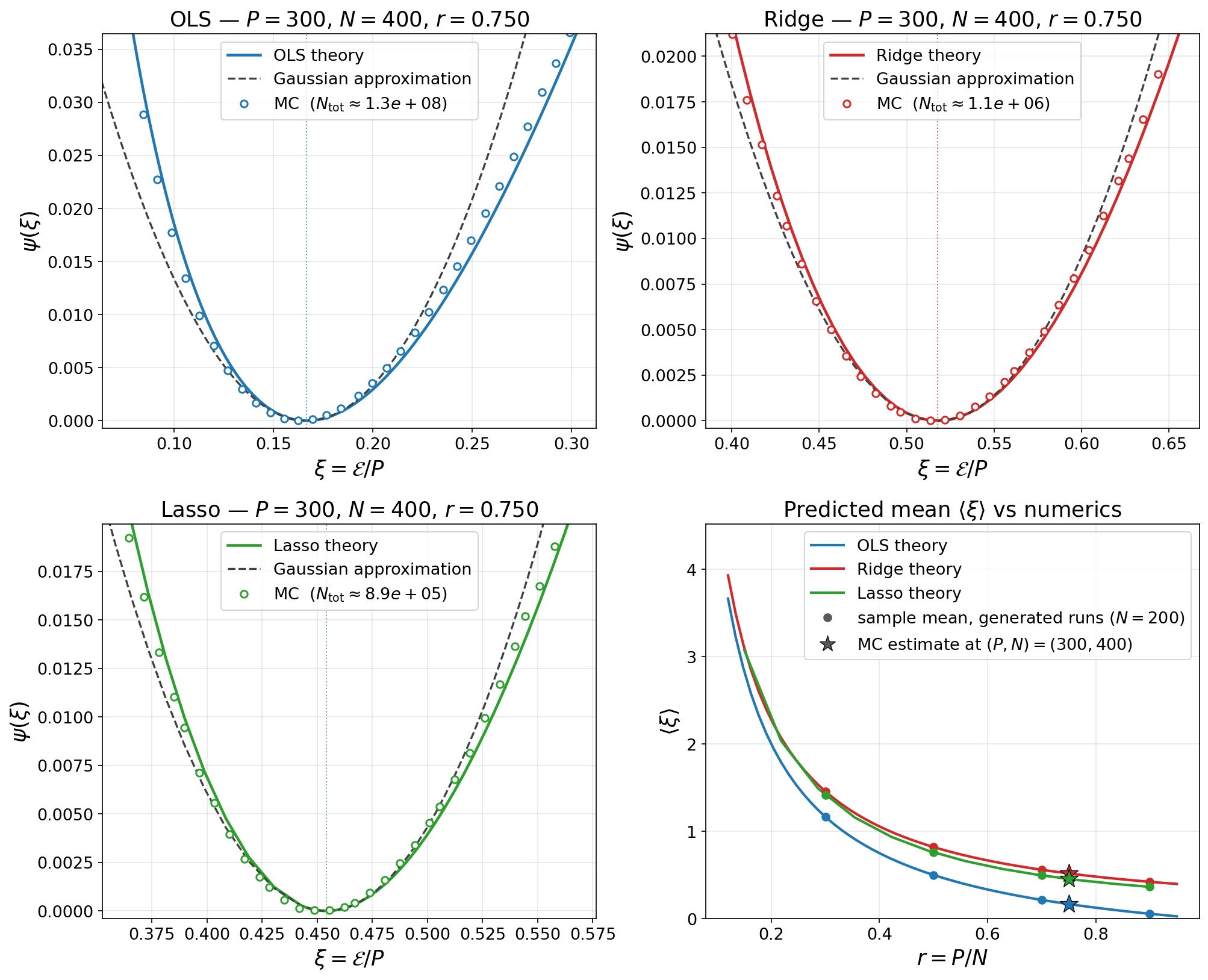}
  \caption{Distribution and mean of the minimal training loss for OLS,
Ridge, and Lasso at $P = 300$, $N = 400$ ($r = P/N = 0.750$), with $p(\beta)=\delta(\beta-1)$.
Top left, top right, and bottom left: large-deviation rate function
$\psi(\xi)$ of the intensive loss $\xi = \mathcal{E}/P$ for OLS, Ridge
($\lambda = 0.5$), and Lasso ($\lambda = 0.5$),
respectively. Solid lines: exact replica prediction (Eq.~\eqref{eq:psixi});
dashed lines: Gaussian approximation about the typical value; open
circles: rate function estimated from
$N_{\mathrm{tot}} = 1.1\times 10^{9}$ (OLS),
$1.1\times 10^{6}$ (Ridge), and $8.9\times 10^{5}$ (Lasso) independent
realizations of the disorder ($X$, $\bm{\epsilon}$, $\bm{\beta}$).
Both theory and simulation show clear deviations from the Gaussian
approximation away from the typical value.
Bottom right: predicted mean loss $\langle \xi \rangle = \Phi'(0)$ vs.\
$r$ for the three models (solid lines), compared with the sample mean
over MC generated realizations at several values of $r$ (open circles, at $N=200$)
and a single Monte Carlo estimate at $P = 300$, $N = 400$ used in the
other three panels (stars).}
\label{fig:distribution}
\end{figure}

The object of interest is the random variable
\begin{equation}
   \mathcal{E}=\min_{\bm w} H_\lambda(\bm w)\ ,\label{mathcalEmindef}
\end{equation}
where the randomness arises from the Gaussian design $\bm x$, the noise $\bm\epsilon$, and the teacher weights $\{\beta_j\}_{j=1}^P$, drawn independently from $p(\beta)$. For
$P,\, N \to \infty$ with their ratio $r = P/N$ fixed, and under certain light-tailedness assumptions on an otherwise general $p(\beta)$, we show that the full distribution of the minimal regularized training loss takes the large deviation form
\begin{equation}
\mathbb{P}(\mathcal{E}= P \xi) \approx {\re}^{-P \psi(\xi)}\ \label{PratePsi}
\end{equation}
in terms of the rate function $\psi(\xi)$: the exponential cost of observing an intensive loss value $\xi=\mathcal{E}/P$ far from its typical value. The rate function has a quadratic behavior near its minimum $\xi^\star$, corresponding to Gaussian typical fluctuations. This result, valid for any separable regularizer, is further specialized to OLS, Ridge and Lasso (see Fig. \ref{fig:distribution}).

Our approach is rooted in the statistical mechanics of disordered systems. We employ
the \emph{zero-temperature replica method}~\cite{Mezard}, a technique developed to
analyze systems with quenched disorder such as spin glasses by treating the
minimal loss as the ground-state energy of a random cost function. The replica
approach in the enhanced formulation put forward in \cite{FyodorovTopo} is able to extract not only the \emph{average} minimal loss, but also the full large-deviation structure in a unified framework.

\textit{Summary of results - }The scaled cumulant generating function $\Phi(s)$ of $\mathcal{E}$, defined as
\begin{equation}
\Phi(s) = - \lim_{P\to\infty} \frac{1}{P} 
\log\langle {\re}^{-s \mathcal{E}} \rangle\ , \label{eq:fullPhis}
\end{equation}
where $\langle\cdot \rangle$ denotes averaging with respect to the Gaussian disorder and the $\beta_j$'s, is given by the general formula
\begin{align}
\nonumber-\Phi(s)
&=
s(\hat q_0\chi+q_0\hat\chi)
+s^2q_0\hat q_0+
\log
\left\langle {\re}^{-s m(\beta,z)}\right\rangle_{z,\beta}\\
&-\frac{1}{2r}
\log
\frac{1+\chi+s(q_0+1)}{1+\chi}\ ,\label{PhisEqSummary}
\end{align}
where
\begin{equation}
    V_\lambda(w)=\hat\chi(\beta-w)^2
-
\sqrt{-2\hat q_0}\,z(\beta-w)
+
g_\lambda(w) ,
\end{equation}
and
$
m(\beta,z)
=
\min_w V_\lambda(w)$,  $w^\star(\beta,z)=\mathrm{argmin} ~V_\lambda(w)$\ .   
Here
\begin{equation}
\left\langle \cdots \right\rangle_{z,\beta}
=
\int \mathrm{D}z\int \dd\beta\,p(\beta)(\cdots)\ ,
\qquad
\mathrm{D}z=
\frac{\dd z}{\sqrt{2\pi}}{\re}^{-z^2/2}\ ,
\end{equation}
provided the integral $\int \dd\beta\,p(\beta)(\cdots)$ is finite. The parameters $\chi,\hat\chi,q_0,\hat q_0$ (all functions of $s$) satisfy the set of coupled nonlinear equations 
\begin{align}
\nonumber\hat\chi
&=
\frac{1}{2r(1+\chi)}\ ,
\\
\nonumber\hat q_0
&=
\frac{1}{s}
\left[
\frac{1}{2r\left[1+\chi+s(q_0+1)\right]}
-
\hat\chi
\right]\ ,
\\
\nonumber q_0
&=
\left\langle
\left(\beta-w^*(\beta,z)\right)^2
\right\rangle_s\ ,
\\
\chi+s q_0
&=
\frac{
\left\langle
z\left(\beta-w^*(\beta,z)\right)
\right\rangle_s
}{
\sqrt{-2\hat q_0}
}\ .\label{eqorderparameter}
\end{align}
In the above, we also defined the tilted expectation
$
\left\langle F\right\rangle_s
=\left\langle F(\beta,z){\re}^{-s m(\beta,z)}\right\rangle_{z,\beta}
/
\left\langle {\re}^{-s m(\beta,z)}\right\rangle_{z,\beta}
$.
From the standard theory of large deviations \cite{touchette}, it follows that Eq. \eqref{PratePsi} holds, with 
\begin{equation}
    \psi(\xi) = \sup_{s} \left\{\Phi(s)-s\xi \right\}\ .
    \label{eq:psixi}
\end{equation}
From $\Phi(s)$ in Eq. \eqref{eq:fullPhis}, all cumulants of the minimal training loss can be computed explicitly to leading order by differentiation. For instance
$    \lim_{P\to\infty} \langle\mathcal{E}\rangle/P=\Phi'(0)$ (see \cite{SM} for explicit expressions). In the OLS case (no regularization), the formulas above collapse for $r<1$ to $\Phi(s)=\frac{1-r}{2r}\log(1+s)$ and the rate function to $ \psi(\xi)
    =
    \xi-\frac{1-r}{2r}
    -\frac{1-r}{2r}
    \log\left(
        \frac{2r\xi}{1-r}
    \right)$, which has a minimum at the known value of the typical optimal training loss per parameter $
    \xi_\star=(1-r)/(2r)$. As $r$ increases from zero, the typical optimal training loss decreases and vanishes at the critical value $r=1$. This coincides with the divergence of the order parameters $q_0$ and $\chi$, which respectively quantify the estimation error and the susceptibility. The simultaneous vanishing of the training loss and divergence of these order parameters signals the onset of overfitting, where the model is able to interpolate the training data while becoming increasingly sensitive to the particular realization of the dataset. Introducing regularization removes the singularity at $r=1$, keeping the order parameters finite.
    Furthermore, the expression \eqref{PhisEqSummary} specialized to the Ridge case for Gaussian $p(\beta)$ is perfectly compatible with the quadratic optimization result in \cite{FyodorovTopo} (see also \cite{vivonotes}) and with further consistency checks involving Wishart matrices \cite{SM}.

    The light-tailedness assumption on $p(\beta)$ is not merely technical. The $P$-speed large-deviation form in Eq. \eqref{PratePsi} requires the tilted single-site average $\langle \mathrm{e}^{-s m(\beta,z)}\rangle_{z,\beta}$ to remain finite over the relevant range of $s$. If $p(\beta)$ does not possess the required exponential moments, this condition may fail, particularly for the upper-loss tail ($s<0$). Rare fluctuations of the optimal regularized cost may then be generated not by a collective $O(P)$ fluctuation of the disorder, but by one or a few exceptionally large teacher coefficients. This is the single-big-jump mechanism characteristic of heavy-tailed sums \cite{ArmendarizLoulakis2011,VezzaniBarkaiBurioni2019}, and can be associated with a \emph{condensation} of a finite fraction of the fluctuation onto a single degree of freedom \cite{SzavitsNossanEvansMajumdar2014}. Depending on the tail of $p(\beta)$, the conventional exponential scaling $\exp[-P\psi(\xi)]$ may therefore break down or be replaced by a slower, tail-dependent large-deviation speed \cite{Lehtomaa2017}. For sufficiently heavy tails, even the $1/\sqrt{P}$ normalization of the teacher signal--and hence the assumed $O(P)$ extensivity of the cost--may need to be reconsidered. This caveat concerns the regularized problems; for OLS, the teacher coefficients can be shifted out of the minimization and the optimal residual loss is independent of $p(\beta)$. A detailed analysis of these anomalous large deviation mechanisms for heavy-tailed $p(\beta)$ is deferred to a separate publication.

\textit{Sketch of the derivation - }We introduce the partition function at inverse temperature $\gamma$
\begin{equation}
 Z(\gamma)=\int_{\mathbb R^P}\dd\bm w\;\exp[-\gamma H_\lambda(\bm w)]\ ,
\label{eq:partition_function}
\end{equation}
so that the minimization problem for large $\gamma$ can be mapped to the following
\begin{equation}
  \mathcal{E}=\min_{\bm w}H_\lambda(\bm w)=   \lim_{\gamma\to\infty} -\frac{1}{\gamma} \log Z(\gamma)\ .
\end{equation}
The cumulant generating function of $\mathcal{E}$ can be written as \cite{FyodorovTopo} 
\begin{equation}
    \langle {\re}^{-s \mathcal{E}} \rangle=\lim_{\gamma\to\infty} \langle Z^{s/\gamma}(\gamma)\rangle\ ,
\end{equation}
where $Z^n(\gamma)$ is the partition function in Eq. \eqref{eq:partition_function} replicated $n$ times, with $n$ initially taken as an integer. The average of the replicated partition function to compute is therefore
\begin{align}
 \nonumber   &\langle Z^n(\gamma)\rangle  =\int \prod_j \dd\beta_j\,p(\beta_j)\int \prod_{a=1}^n \dd\bm w_a\prod_{i,j} \mathrm{D} x_j^{(i)}  \prod_{i} \mathrm{D}\epsilon^{(i)} \\
    &\times \exp\left[{-\gamma \frac{1}{2}\sum_{i,a} \left(y^{(i)} - \sum_{j}\frac{w_{ja}}{\sqrt{P}} x_j^{(i)}\right)^2 - \gamma \sum_{j,a}g_{\lambda}(w_{ja})}\right]\ .
\end{align}
Performing the Gaussian $\bm x$ and $\bm\epsilon$ integrals with the help of a standard Hubbard-Stratonovich transformation to handle the squares, we get to $\langle Z^n(\gamma)\rangle_{\bm x,\bm\epsilon}=$
\begin{align}
 \nonumber   &    =\Big\langle {\rm e}^{P\sum_{ab}\hat Q_{ab}Q_{ab}-\sum_{abj}\hat Q_{ab} (\beta_j-w_{ja})(\beta_j-w_{jb}) -\gamma\sum_{ja}g_\lambda(w_{ja})}\\
 &\times\int\prod_{ia} \dd \phi^{{(i)}}_{a}  {\rm e}^{-\frac{1}{2} \sum_{ia} \phi^{(i)}_a(\delta_{ab}+\gamma (1+Q_{ab}))\phi^{(i)}_b }\Big\rangle_{Q,\hat Q,\bm w,\bm\beta}\ ,
\end{align}
where we have introduced the matrix order parameter $Q_{ab}=(1/P)\sum_j (\beta_j-w_{ja})(\beta_j-w_{jb})$ and enforced its definition via the conjugate matrix $\hat Q$. Adopting a standard replica symmetric ansatz 
\be
Q =
\begin{cases}
q_0 + \chi & \text{if } a = b \\
q_0 & \text{if } a \ne b
\end{cases}
\quad\hat{Q} =
\begin{cases}
\hat{q}_0 + \hat{\chi} & \text{if } a = b \\
\hat{q}_0 & \text{if } a \ne b\ ,
\end{cases}
\ee
and solving in turn the Gaussian integrals in $\bm\phi_a$ and in $\bm w$ with the help of a further Hubbard-Stratonovich linearization with Gaussian parameter $z$, we get after the rescaling $\chi\to\chi/\gamma,
\hat q_0\to\gamma^2\hat q_0,
\hat\chi\to\gamma\hat\chi$ and the limit $\gamma\to\infty$ the expression
\begin{equation}
\langle Z^{s/\gamma}(\gamma)\rangle
\sim
\int \dd q_0\,\dd\chi\,\dd\hat q_0\,\dd\hat\chi\,
\exp\left[
P\,\mathcal{S}_s(q_0,\chi,\hat q_0,\hat\chi)
\right]\ ,\label{ZnFDamenable}
\end{equation}
where the finite-\(s\) action is

\begin{align}
&\mathcal{S}_{s}(q_0,\chi,\hat q_0,\hat\chi)
=
s(\hat q_0\chi+q_0\hat\chi)
+
s^2q_0\hat q_0\\
&-\frac{1}{2r}
\log
\left(
\frac{1+\chi+s(q_0+1)}{1+\chi}
\right)
+
\log
\left\langle
{\re}^{-s\,m(\beta,z)}
\right\rangle_{z,\beta}.\label{eq:Psi_s_general}
\end{align}
The form in Eq. \eqref{ZnFDamenable} is amenable to a saddle-point evaluation for large $P$, which eventually leads to the saddle point equations \eqref{eqorderparameter} and the scaled cumulant generating function in \eqref{PhisEqSummary} (see \cite{SM} for the extended derivation).\\

\textit{Conclusions and Outlook - } We have shown that the optimized cost in high-dimensional linear regression has a nontrivial large-deviation structure. Using a zero-temperature replica construction, we derived the scaled cumulant generating function for a broad class of separable regularizers, including Ordinary Least Squares, Ridge, and Lasso, giving access to both the typical value and exponentially rare sample-to-sample fluctuations.

These results complement theories of typical high-dimensional behavior by addressing the reliability of inference from a single dataset. The large-deviation rate function quantifies how representative an atypically good or bad optimum is, while regularization reshapes not only the typical solution but the entire rare-event structure. OLS provides an exactly solvable benchmark, whereas Ridge and Lasso illustrate how quadratic and sparse penalties modify these tails. 

More broadly, the framework suggests treating reliability in high-dimensional inference as a problem of rare events. Natural extensions include correlated designs, non-Gaussian noise, fixed signal realizations, and generalized linear models. Heavy-tailed signal distributions are particularly interesting: when exponential integrability fails, the conventional $P$-speed large-deviation principle may change, with rare fluctuations potentially dominated by a condensation or single-big-jump mechanism. Such extensions may clarify when an apparently successful fit reflects a stable structure rather than an atypical realization of the disorder.

%%%%%%%%%%%%%%%%%%%%%%%%%%%%%%%%%%references
% --- the next line keeps `References' out of the Supplement's table of contents
\makeatletter\let\SMsaved@addcl\addcontentsline\renewcommand\addcontentsline[3]{}\makeatother

\makeatletter\let\addcontentsline\SMsaved@addcl\makeatother

% ============================================================================
%                          SUPPLEMENTAL MATERIAL
% ============================================================================
\clearpage
\onecolumngrid

% --- number the sections (a PRL Letter numbers none) and restart counters ---
\setcounter{secnumdepth}{3}
\setcounter{section}{0}
\setcounter{equation}{0}
\setcounter{figure}{0}
\setcounter{table}{0}
\renewcommand{\thesection}{S\arabic{section}}
\renewcommand{\theequation}{S\arabic{equation}}
\renewcommand{\thefigure}{S\arabic{figure}}
\renewcommand{\thetable}{S\arabic{table}}
% --- keep hyperref anchors unique across the two parts ---------------------
\renewcommand{\theHsection}{SM.\arabic{section}}
\renewcommand{\theHequation}{SM.\arabic{equation}}
\renewcommand{\theHfigure}{SM.\arabic{figure}}
\renewcommand{\theHtable}{SM.\arabic{table}}

\begin{center}
{\large\textbf{Supplemental Material:\\[2pt] Large deviations for linear regressions}}\\[10pt]
Silvia Bartolucci,$^{1}$ Fabio Caccioli,$^{1}$ Francesco Caravelli,$^{2}$ and Pierpaolo Vivo$^{3}$\\[6pt]
{\small\itshape
$^{1}$Department of Computer Science, University College London, WC1E 6EA, London (UK)\\
$^{2}$T-Division (T-4), Los Alamos National Laboratory, Los Alamos NM, 87544 (USA)\\
$^{3}$Department of Mathematics, King's College London, WC2R 2LS, London (UK)}
\end{center}

\vspace{1em}
\tableofcontents
\vspace{1em}

\section{Partition function for the minimal loss}
We study the regularized objective function [cf.\ Eq.~(3) of the main text]
\begin{equation}
 H_\lambda(\bm w)
 =\frac12\sum_{i=1}^N\left(y^{(i)}-\frac1{\sqrt P}\sum_{j=1}^P w_jx_j^{(i)}\right)^2
 +\sum_{j=1}^P g_\lambda(w_j)\ ,
\label{eq:general_objective}
\end{equation}
where $g_\lambda$ is a general separable regularizer.  The factor $1/\sqrt{P}$ has been introduced so that, with weights of order $1$, $\sum_{j=1}^P\frac{w_j}{\sqrt{P}} x_j^{(i)}$ is of order $1$ like $y^{(i)}$, and the cost function is extensive. The goal is to compute the statistics of 
$\mathcal{E}=\min_{\bm w}H_\lambda(\bm w)$.
We introduce the partition function at inverse temperature $\gamma$
\begin{equation}
 Z(\gamma)=\int_{\mathbb R^P}\dd\bm w\;\exp[-\gamma H_\lambda(\bm w)]\ ,
\label{sm:eq:partition_function}
\end{equation}

so that the minimization problem for large $\gamma$ can be mapped to the following
\begin{equation}
  \min_{\bm w}H_\lambda(\bm w)=   \lim_{\gamma\to\infty} -\frac{1}{\gamma} \log Z(\gamma)\ .
\end{equation}
\section{Replica calculation for average minimal loss}
We compute the averages over the quenched disorder, namely: 
\begin{equation}\langle\mathcal{E}\rangle_{\bm x,\bm\epsilon,\bm\beta}=\langle[\min_{\bm w}H_\lambda(\bm w)]\rangle_{\bm x,\bm\epsilon,\bm\beta}=   \lim_{\gamma\to\infty} -\frac{1}{\gamma} \langle\left[\log Z(\gamma)\right]\rangle_{\bm x,\bm\epsilon,\bm\beta}\ .\label{eq:Easlog}
\end{equation}
These averages can be computed using the replica identity:
\begin{equation}
 \langle\log Z(\gamma)\rangle_{\bm x,\bm\epsilon,\bm\beta}=\lim_{n\to0}\frac{1}{n}\log \langle Z^n(\gamma)\rangle\ ,
\label{eq:replica_identity}
\end{equation}
where $n$ is initially treated as an integer, and $\langle (\cdots)\rangle$ is a shorthand for $\langle (\cdots)\rangle_{\bm x,\bm\epsilon,\bm\beta}$.

Let  us write the replicated partition function explicitly, introducing Gaussian integration over the quenched variables $\bm x,\bm \epsilon$ as  
\begin{align}
\dd\mu(\bm x)&{\propto} \prod_{i,j} \dd x_j^{(i)} \re^{-\frac{1}{2}\sum_{i,j} (x_j^{(i)})^2}\ , \nonumber \\
\dd\mu(\bm \epsilon)&{\propto} \prod_{i} \dd\epsilon^{(i)} \re^{-\frac{1}{2}\sum_{i} (\epsilon^{(i)})^2}\ ,\nonumber
\end{align}
and the shorthand
\begin{equation}
    \int \{\dd \}=\int \prod_j \dd\beta_j\,p(\beta_j)\int \prod_a \dd\bm w_a~{\dd\mu(\bm x)} \ .
\end{equation}

Hence, we have
\begin{align}
    & \langle Z^n(\gamma)\rangle  =\int \{\dd \}~\dd\mu(\bm\epsilon)~{\rm e}^{-\gamma \frac{1}{2}\sum_{i,a} \left(y^{(i)} - \sum_{j}\frac{w_{ja}}{\sqrt{P}} x_j^{(i)}\right)^2 - \gamma \sum_{j,a}g_{\lambda}(w_{ja})}\label{eq:step0}\\
    &\approx\int \{\dd \}\int \prod_{ia}\dd\phi^{{(i)}}_{a}~\dd\mu(\bm\epsilon)~\re^{-\mathrm{i}\sqrt{\gamma}\sum_{i,a} \left(y^{(i)} - \sum_{j}\frac{w_{ja}}{\sqrt{P}} x_j^{(i)}\right)\phi^{{(i)}}_{a}-\frac{1}{2}\sum_{i,a}{\phi^{{(i)}}_{a}}^2 - \gamma \sum_{j,a}g_{\lambda}(w_{ja})  }\\
    &\approx\int \{\dd \}\int \prod_{ia}\dd \phi^{{(i)}}_{a}~\prod_{i} \dd\epsilon^{(i)}~\re^{-\mathrm{i}\sqrt{\gamma}\sum_{i,a} \left(\sum_j \frac{\beta_j}{\sqrt{P}} x_j^{(i)}+\epsilon^{(i)}- \sum_{j}\frac{w_{ja}}{\sqrt{P}} x_j^{(i)}\right)\phi^{{(i)}}_{a}-\frac{1}{2}\sum_{i,a}{\phi^{{(i)}}_{a}}^2-\frac{1}{2}\sum_{i}\epsilon^{(i)2} - \gamma \sum_{j,a}g_{\lambda}(w_{ja})}\\
    &\approx\int \{\dd \}\int \prod_{ia}\dd\phi^{{(i)}}_{a}~~\re^{-\mathrm{i}\sqrt{\gamma}\sum_{i,a} \left(\sum_j \frac{\beta_j}{\sqrt{P}} x_j^{(i)}- \sum_{j}\frac{w_{ja}}{\sqrt{P}} x_j^{(i)}\right)\phi^{{(i)}}_{a}-\frac{1}{2}\sum_{i,a}{\phi^{{(i)}}_{a}}^2-\frac{\gamma}{2}\sum_{i}(\sum_a \phi^{{(i)}}_{a})^2- \gamma \sum_{j,a}g_{\lambda}(w_{ja})}\\
     &=\int \{\dd \}\int \prod_{ia}\dd\phi^{{(i)}}_{a}~\re^{-\mathrm{i}\sqrt{\gamma}\sum_{i,a} \left(\sum_j \frac{\beta_j}{\sqrt{P}} x_j^{(i)}- \sum_{j}\frac{w_{ja}}{\sqrt{P}} x_j^{(i)}\right)\phi^{{(i)}}_{a}-\frac{1}{2}\sum_{i,a}{\phi^{{(i)}}_{a}}^2-\frac{\gamma}{2}\sum_{i}\sum_{a,b} \phi^{{(i)}}_{a}\phi^{{(i)}}_{b}- \gamma \sum_{j,a}g_{\lambda}(w_{ja})}\ ,\label{eq:step5}
\end{align}
where we have first applied a Hubbard-Stratonovich transformation, then we made the dependence of $\bm\epsilon$ explicit, finally we integrated over $\bm{\epsilon}$.
Integrating over the $\bm x$ Gaussian variables and expanding we get for the $\bm\phi$- and $\bm w$-integrals:
 \bea
\nonumber &\int\prod_{ia} \dd \phi^{{(i)}}_{a}\prod_{ja} \dd w_{ja} ~\re^{-\frac{1}{2}\sum_{ia}{\phi^{(i)}_a}^2-\frac{\gamma}{2}\sum_{iab}\phi^{(i)}_a\phi^{(i)}_b-\frac{\gamma}{2}\sum_{ij}\left(\sum_a \left(\frac{\beta_j}{\sqrt{P}}- \frac{w_{ja}}{\sqrt{P}}\right)\phi^{(i)}_a\right)^2- \gamma \sum_{j,a}g_{\lambda}(w_{ja})}=\\
   &=\int\prod_{ia} \dd \phi^{{(i)}}_{a}\prod_{ja} \dd w_{ja} ~\re^{-\frac{1}{2}\sum_{ia}{\phi^{(i)}_a}^2-\frac{\gamma}{2}\sum_{iab}\phi^{(i)}_a\phi^{(i)}_b-\frac{\gamma}{2}\sum_{ij}\sum_{ab}\phi^{(i)}_a\phi^{(i)}_b\frac{(\beta_j-w_{ja})(\beta_j-w_{jb})}{P}- \gamma \sum_{j,a}g_{\lambda}(w_{ja})}\ .\quad\quad 
\eea
 
We also introduce the overlap matrices $Q_{ab}=(1/P)\sum_j (\beta_j-w_{ja})(\beta_j-w_{jb})$ and enforce their definition via delta integrals: 
\begin{eqnarray}
    1&&=\int\prod_{ab}\dd Q_{ab}\delta(PQ_{ab}-\sum_j (\beta_j-w_{ja})(\beta_j-w_{jb})) \label{eq:defQabinline}\\
    &&=\frac{1}{(2\pi)^{n^2}}\int_{-i\infty}^{i\infty} \prod_{ab}\dd\hat Q_{ab}\int\prod_{ab}\dd Q_{ab} ~{\rm e}^{\sum_{ab} \hat Q_{ab}(PQ_{ab}-\sum_j (\beta_j-w_{ja})(\beta_j-w_{jb}))}\\
    &&\propto\int_{-i\infty}^{i\infty} \prod_{ab}\dd\hat Q_{ab}\int\prod_{ab}\dd Q_{ab} ~{\rm e}^{\sum_{ab} \hat Q_{ab}(PQ_{ab}-\sum_j (\beta_j-w_{ja})(\beta_j-w_{jb}))}\ ,
\end{eqnarray}
where we neglected an overall constant factor in the exponent and irrelevant to the later saddle point for large $P$. The average partition  function can be then rewritten as:
\begin{align}
\nonumber   \langle Z^n(\gamma) \rangle &=\int \prod_j \dd\beta_j p(\beta_j)\int\prod_{ab}\dd Q_{ab}\dd\hat Q_{ab}\int\prod_{ia} \dd \phi^{{(i)}}_{a}\prod_{ja} \dd w_{ja}\times\\
    &\times {\rm e}^{-\frac{1}{2}\sum_{ia}\phi_a^{(i)2}-\frac{\gamma}{2}\sum_{iab}\phi_a^{(i)}\phi_b^{(i)}-\frac{\gamma}{2}\sum_{iab}\phi_a^{(i)}Q_{ab}\phi_b^{(i)}+P\sum_{ab}\hat Q_{ab}Q_{ab}-\sum_{ab}\hat Q_{ab}\sum_j (\beta_j-w_{ja})(\beta_j-w_{jb}) - \gamma \sum_{ja}g_{\lambda}(w_{ja})} = \nonumber\\
&  =\int \prod_j \dd\beta_j p(\beta_j)\int\prod_{ab}\dd Q_{ab}\dd \hat Q_{ab}\prod_{ja} \dd w_{ja}\times\nonumber\\&\times {\rm e}^{P\sum_{ab}\hat Q_{ab}Q_{ab}-\sum_{ab}\hat Q_{ab}\sum_j (\beta_j-w_{ja})(\beta_j-w_{jb}) -\gamma\sum_{ja}g_\lambda(w_{ja})}  \int\prod_{ia} \dd \phi^{{(i)}}_{a}  {\rm e}^{-\frac{1}{2} \sum_{ia} \phi^{(i)}_a(\delta_{ab}+\gamma (1+Q_{ab}))\phi^{(i)}_b } \ ,
\end{align}
where we rearranged the $\bm{\phi}$'s terms into a square. 
We will now compute the different parts of the average partition function separately under a replica symmetric ansatz \footnote{{We can write $Q=a I+b J$, and $\hat Q=c I + dJ$, where $I$ is the identity and $J$ is the matrix with zero diagonal and 1's on the off-diagonal. Above, $a=q_0+\chi$, $b=q_0$, $c=\hat q_0+\hat \chi$ and $d=\hat q_0$. Note that we have $\text{trace}({I})=n$, $\text{trace}({J})=0$ and $\text{trace}(J^2)=n(n-1)$, since $J^2$ is the matrix with  $n-1$ on the diagonal. }}, where
\be
Q_{ab} =
\begin{cases}
q_0 + \chi & \text{if } a = b \ , \\
q_0 & \text{if } a \ne b \ ,
\end{cases}\label{QabRSassumptiondef}
\ee
\be
\hat{Q}_{ab} =
\begin{cases}
\hat{q}_0 + \hat{\chi} & \text{if } a = b \ , \\
\hat{q}_0 & \text{if } a \ne b\ .
\end{cases}
\ee
\paragraph{$\star$ G-term and $\phi$ integration}
We also perform the Gaussian integral over the $\bm \phi$ variables, which gives
\be
 {\int{\prod_{ia} \dd\phi^{{(i)}}_{a}  \re^{-\frac{1}{2} \sum_{iab }\phi^{(i)}_a\left( \gamma+\delta_{ab}+\gamma Q_{ab}\right)\phi^{(i)}_b}} = \left(\frac{(2\pi)^n}{{\rm det} \left(G\right)}\right)^{N/2}},
\ee
where $G$ is the matrix with elements $G_{ab}=\left(\gamma+\delta_{ab}+\gamma Q_{ab}\right)$. The matrix has the following form:
\begin{equation}
G =
\begin{pmatrix}
1+\gamma(1+q_0+\chi) & \gamma(1+q_0) & \cdots & \gamma(1+q_0) \\
\gamma(1+q_0) & 1+\gamma(1+q_0+\chi) & \cdots & \gamma(1+q_0) \\
\vdots & \vdots & \ddots & \vdots \\
\gamma(1+q_0) & \gamma(1+q_0) & \cdots & 1+\gamma(1+q_0+\chi)
\end{pmatrix}\ .
\end{equation}
The determinant of $G$ can easily be written in terms of the product of its eigenvalues. In the replica symmetric ansatz, the matrix $G$ has in fact one eigenvalue associated with an eigenvector of entries all ones, and one eigenvalue, of multiplicity $n-1$, associated with eigenvectors with two subsequent entries equal to $+1$ and $-1$,  all other entries being zero. The eigenvalues read
\bea
\lambda_\parallel &=& \gamma q_0+\gamma\chi+\gamma(n-1) q_0+1+n\gamma = \gamma\chi+\gamma n q_0+1+\gamma n\ ,\\
\lambda_\perp&=& \gamma q_0+\gamma\chi+\gamma+1-\gamma q_0-\gamma =\gamma\chi+1\ ,
\eea
with multiplicity respectively $1$ and $n-1$. Hence, 
\be
{{\rm det}(G)=( \gamma\chi+\gamma n q_0+1+\gamma n)(\gamma\chi+1)^{n-1}}\ .
\label{eq:detG_expansion_general}
\ee
Once exponentiated, and neglecting constant terms, this term contributes as
\bea
-\frac{N}{2} \left[\log (\gamma\chi+n q_0\gamma+1+n\gamma)+(n-1)\log(1+\gamma\chi)\right]\ ,\\
=-\frac{N}{2} \left[\log \left(1+\frac{n\gamma q_0+n\gamma}{1+\gamma\chi}\right)+n\log(1+\gamma\chi)\right]\ ,\\
\label{eq:free1}\approx -\frac{N}{2} \left[\frac{n\gamma q_0+n\gamma}{1+\gamma\chi}+n\log(1+\gamma\chi)\right]\ ,
\eea
where in the last step we have neglected all terms of order higher than $n$ in the expansion of the logarithm.
\paragraph{$\star$ Overlap term}
For the element-wise product of the matrix of overlap and its conjugate, we get the following term in the exponent \footnote{This can be shown algebraically. We can in fact write $\sum_{ab} Q_{ab} \hat Q_{ab}=\text{trace}(Q\hat Q)$ since these are symmetric matrices, and we have $Q\hat Q=ac I+(ad+bc) J+ bd J^2$. Using the fact that $\text{trace}(I)=n, \text{trace}(J)=0$ and $\text{trace}(J^2)=n(n-1)$, we obtain $\text{trace}(Q\hat Q)=n ac +bd n (n-1)$. Keeping only the order linear in $n$, we have $\text{trace}(Q\hat Q)=n(ac-bd)+O(n^2)\approx n((q_0+\chi)(\hat q_0+\hat \chi)-q_0\hat q_0)$ which gives the result above.}
\begin{align}
\nonumber P\sum_{ab}\hat Q_{ab}Q_{ab} &= P\left[n(q_0+\chi)(\hat q_0+\hat\chi)+n(n-1)q_0\hat q_0\right] = P\left[n (\hat q_0\chi + q_0 \hat{\chi} + \hat{\chi}\chi)  + n^2 q_0 \hat{q_0}\right] \\
   &  \approx_{n\rightarrow 0} P n (\hat q_0\chi+q_0\hat\chi+\chi\hat\chi)\ , 
     \label{eq:Pterm}
\end{align}
where again we only kept terms of order $n$.
\paragraph{$\star$ $w$-integration}
{We now turn to the part of the partition function that depends on the $w$'s. 

\bea
&&
\int \prod_j \dd\beta_j\,p(\beta_j)
\int \prod_{ja} \dd w_{ja}\,
\exp\left[
-\sum_{j,a,b}\hat Q_{ab}
(\beta_j-w_{ja})(\beta_j-w_{jb})
-\gamma\sum_{j,a}g_\lambda(w_{ja})
\right]=
\nonumber\\
&&=
\prod_{j=1}^{P}
\left(
\int \dd\beta_j\,p(\beta_j)
\int \prod_a \dd w_a\,
\exp\left[
-\sum_{a,b}\hat Q_{ab}
(\beta_j-w_a)(\beta_j-w_b)
-\gamma\sum_a g_\lambda(w_a)
\right]
\right)\nonumber\\
&&=
\prod_{j=1}^{P}
\left(
\int \dd\beta_j\,p(\beta_j)
\int \prod_a \dd w_a\,
\exp\left[
-\hat\chi\sum_a(\beta_j-w_a)^2
-\hat q_0
\left(
\sum_a(\beta_j-w_a)
\right)^2
-\gamma\sum_a g_\lambda(w_a)
\right]
\right)
\nonumber\\
&&=
\prod_{j=1}^{P}
\left(
\int \underbrace{\dd z~{\rm e}^{-\frac{z^2}{2}}}_{\mathrm{D}z}
\int \prod_a \dd w_a\,
\exp\left[
-\hat\chi\sum_a(\beta_j-w_a)^2
+\sqrt{-2\hat q_0}\,z
\sum_a(\beta_j-w_a)
-\gamma\sum_a g_\lambda(w_a)
\right]
\right)\quad\quad
\label{eq:z_general}\\
&&=
\left(
\int \dd \beta\,p(\beta)
\left\langle
\left[
\int \dd w\,
\exp\left(
-\hat\chi(\beta-w)^2
+\sqrt{-2\hat q_0}\,z(\beta-w)
-\gamma g_\lambda(w)
\right)
\right]^n
\right\rangle_z
\right)^P \ , \label{eq:potenzialegeneral}
\eea
where in the last step we have denoted by $\langle\cdot\rangle_z$ the average over the Gaussian variable $z$, which we introduced through a Hubbard-Stratonovich transformation in \eqref{eq:z_general}. 
}

We now collect the three contributions obtained above: the Gaussian integral over the auxiliary fields $\bm\phi$, the overlap constraint, and the single-site
integral over the replicated weights.

Putting together
\eqref{eq:detG_expansion_general}, \eqref{eq:Pterm},
and \eqref{eq:potenzialegeneral}, we obtain the replicated partition
function in the form
\begin{align}
 \langle Z^n(\gamma)\rangle
&\propto
\int \dd q_0\,\dd\chi\,\dd\hat q_0\,\dd\hat\chi\,
\exp\Bigg\{
-\frac{N}{2}
\left[
\frac{n\gamma(q_0+1)}{1+\gamma\chi}
+
n\log(1+\gamma\chi)
\right]
+
Pn(\hat q_0\chi+q_0\hat\chi+\chi\hat\chi)
\nonumber\\
&\qquad\qquad
+
Pn
\left\langle
\log
\left[
\int \dd w\,
\exp\left(
-\hat\chi(\beta-w)^2
+
\sqrt{-2\hat q_0}\,z(\beta-w)
-\gamma g_\lambda(w)
\right)
\right]
\right\rangle_{z,\beta}
\Bigg\}\ ,
\label{eq:Zn_general_regulariser}
\end{align}
where
\begin{equation}
\langle F(z,\beta)\rangle_{z,\beta}
=
\int \mathrm{D}z\int \dd\beta\,p(\beta)\,F(z,\beta)\ .
\end{equation}

In \eqref{eq:potenzialegeneral}, we have used the following chain
\begin{align}
\nonumber    \left(
\int \dd \beta\,p(\beta)
\left\langle
\left[\cdots\right]^n
\right\rangle_z
\right)^P &\approx\exp\left\{P\log\int\dd\beta p(\beta)(1+n\log [\cdots])\right\}\\
&=\exp\left\{P\log\left(1+n\int\dd\beta p(\beta)\log [\cdots]\right)\right\}\approx \exp\left\{Pn\langle\log[\cdots]\rangle_{z,\beta}\right\}\ ,
\end{align}
valid since $\int\dd\beta p(\beta)=1$ by normalization.

Introducing \(r=P/N\) and performing the zero-temperature scaling
\begin{equation}
\chi\to\frac{\chi}{\gamma}\ ,
\qquad
\hat q_0\to\gamma^2\hat q_0\ ,
\qquad
\hat\chi\to\gamma\hat\chi\ ,
\label{eq:zero_temp_rescaling_general}
\end{equation}
we can rewrite
\begin{align}
 \langle Z^n(\gamma)\rangle
&=
\int \dd q_0\,\dd\chi\,\dd\hat q_0\,\dd\hat\chi\,
\exp\left[
-nP\gamma\,
f_\gamma(q_0,\chi,\hat q_0,\hat\chi)
\right]\ ,
\label{eq:Zn_fgamma_general}
\end{align}
where
\begin{align}
f_\gamma(q_0,\chi,\hat q_0,\hat\chi)
&=
\frac{q_0+1}{2r(1+\chi)}
-\hat q_0\chi
-q_0\hat\chi
-\frac{\chi\hat\chi}{\gamma}
+
\frac{1}{2r\gamma}\log(1+\chi)
\nonumber\\
&\quad
-\frac{1}{\gamma}
\left\langle
\log
\int \dd w\,
\exp\left[
-\gamma
\left(
\hat\chi(\beta-w)^2
-
\sqrt{-2\hat q_0}\,z(\beta-w)
+
g_\lambda(w)
\right)
\right]
\right\rangle_{z,\beta}.
\label{eq:fgamma_general_regulariser}
\end{align}

In the limit \(\gamma\to\infty\), Laplace's method yields for the $w$-integral
\begin{equation}
-\frac{1}{\gamma}
\log
\int \dd w\,
{\re}^{-\gamma V(w;\beta,z)}
\longrightarrow
\min_w V(w;\beta,z)\ ,
\end{equation}
with
\begin{equation}
V(w;\beta,z)
=
\hat\chi(\beta-w)^2
-
\sqrt{-2\hat q_0}\,z(\beta-w)
+
g_\lambda(w)\ .
\label{eq:single_site_potential_general}
\end{equation}
Therefore the zero-temperature free energy is
\begin{equation}
f(q_0,\chi,\hat q_0,\hat\chi)
=
\frac{q_0+1}{2r(1+\chi)}
-\hat q_0\chi
-q_0\hat\chi
+
\left\langle
\min_w
\left[
\hat\chi(\beta-w)^2
-
\sqrt{-2\hat q_0}\,z(\beta-w)
+
g_\lambda(w)
\right]
\right\rangle_{z,\beta}.
\\
\label{eq:f_general_regulariser}
\end{equation}

Let
\begin{equation}
w^*(\beta,z)
=
\operatorname*{argmin}_{w}
\left[
\hat\chi(\beta-w)^2
-
\sqrt{-2\hat q_0}\,z(\beta-w)
+
g_\lambda(w)
\right].
\label{eq:wstar_general}
\end{equation}
Then the saddle-point equations are
\begin{align}
\frac{\partial f}{\partial q_0}=0
&\quad\Rightarrow\quad
{
\hat\chi
=
\frac{1}{2r(1+\chi)}
}\ ,
\label{eq:sp_general_q0}
\\
\frac{\partial f}{\partial \chi}=0
&\quad\Rightarrow\quad
{
\hat q_0
=
-\frac{q_0+1}{2r(1+\chi)^2}
}\ ,
\label{eq:sp_general_chi}
\\
\frac{\partial f}{\partial \hat\chi}=0
&\quad\Rightarrow\quad
{
q_0
=
\left\langle
(\beta-w^*)^2
\right\rangle_{z,\beta}
}\ ,
\label{eq:sp_general_chihat}
\\
\frac{\partial f}{\partial \hat q_0}=0
&\quad\Rightarrow\quad
{
\chi
=
\frac{
\left\langle
z(\beta-w^*)
\right\rangle_{z,\beta}
}{
\sqrt{-2\hat q_0}
}
}\ .
\label{eq:sp_general_qhat}
\end{align}

Equations
\eqref{eq:f_general_regulariser}--\eqref{eq:sp_general_qhat}
are the general replica-symmetric zero-temperature equations for a separable
regularizer \(g_\lambda(w)\). The specific choice of regularizer only enters
through the scalar minimization problem defining \(w^*(\beta,z)\).

We now proceed to specializing the saddle point equations and the free energy to specific forms of regularizer or no regularizer to recover the OLS case. Recalling Eq. \eqref{eq:Easlog} and \eqref{eq:replica_identity}, it follows that $\lim_{P\to\infty} \langle\mathcal{E}\rangle/P= f(q_0, \chi,\hat{q_{0}}, \hat{\chi})$ evaluated at the saddle point solutions.

\subsection{OLS (no regularizer)}
In the OLS case, the regularizer is absent (i.e., $g_\lambda(w) =0$). Therefore,  the free energy simplifies to

\be
f(q_0,\chi,\hat q_0,\hat\chi)
=
\frac{q_0+1}{2r(1+\chi)}
-\hat q_0\chi
-q_0\hat\chi
+
\left\langle
\min_w V(w;\beta,z)
\right\rangle_{z,\beta}\ ,
\ee
with 
\be
V(w;\beta,z)
=
\hat\chi(\beta-w)^2
-
\sqrt{-2\hat q_0}\,z(\beta-w)\ .
\ee

The saddle point equations are as follows

\bea
\frac{1}{2r(1+\chi)}-\hat\chi&=&0\ ,
\\
-\hat q_0-\frac{1+q_0}{2r(1+\chi)^2}&=&0\ ,
\\
-q_0+\left\langle (\beta-w^*)^2\right\rangle_{z,\beta}&=&0\ ,
\\
-\chi+\frac{\left\langle z(\beta-w^*)\right\rangle_{z,\beta}}
{\sqrt{-2\hat q_0}}&=&0\ ,
\eea
where $w^*
=
\beta
-
\frac{\sqrt{-2\hat q_0}\,z}{2\hat\chi}
$ is the value of $w$ corresponding to the minimum of the potential $V(w)$.
The solutions of the above system of equations are 
\bea
q_0&=&\frac{r}{1-r}\  , \label{eq:q0OLS}\\
\chi&=&\frac{r}{1-r}\ ,\\
\hat q_0&=&-\frac{1-r}{2r}\ ,\\
\hat\chi&=&\frac{1-r}{2r}\ .
\eea
Inserting the saddle point solutions into the free energy definition, we can compute the minimum of the free energy as
\bea
f(q_0,\chi,\hat q_0,\hat\chi) = \frac{1-r}{2r}\ .\label{minfstarnolambda}
\eea

In Fig. 2 of the main text (bottom right panel) we show full agreement between these formulae and numerical simulations.

We also note that, given the definition of our objective function as one half of the residual sum of squares, and taking into account that $f$ is a free energy per feature, we have the following relationship between the residual sum of squares and the minimum of the free energy:
\be
\mathrm{RSS}/N = 2rf(q_0,\chi,\hat q_0,\hat\chi)=1-r\ .
\ee
The residual sum of squares is a decreasing function of $r$, and it becomes $0$ when $r=1$. This signals the occurrence of overfitting.
In the replica symmetric ansatz, when $\gamma\to\infty$ different replicas of the system end up in the same minimum of the free energy, which is why the order parameter $\chi$ was scaled with $\gamma^{-1}$ earlier. This implies that the order parameter $q_0$ is equal to $q_0=\frac{1}{P}\sum_j \tilde w_j^2$, and is therefore associated with the average error in the estimation of the model parameters. The divergence of $q_0$ for $r=1$ is therefore another sign of overfitting: as $r$ approaches $1$ the solution of the linear regression task is characterized  by a vanishing in-sample error and a diverging estimation error.

\subsection{Ridge regularizer}

For the Ridge regression we consider the regularizer of the form
\begin{equation}
g_\lambda(w)
=
\frac{\lambda}{2}w^2\ .
\label{eq:ridge_regulariser}
\end{equation}
Therefore, the single-site potential simplifies to
\begin{equation}
V_{\mathrm{Ridge}}(w;\beta,z)
=
\hchi(\beta-w)^2
-
\sqrt{-2\hq}\,z(\beta-w)
+
\frac{\lambda}{2}w^2
\label{eq:ridge_potential}
\end{equation}
with minimizer
\begin{equation}
{
w^*_{\mathrm{Ridge}}(\beta,z)
=
\frac{
2\hchi\beta
-
\sqrt{-2\hq}\,z
}{
2\hchi+\lambda
}\ 
} \ .
\label{eq:ridge_wstar}
\end{equation}

Substituting \eqref{eq:ridge_wstar} into the general saddle-point equations (Equations
\eqref{eq:f_general_regulariser}--\eqref{eq:sp_general_qhat}) gives
\begin{align}
\hchi
&=
\frac{1}{2r(1+\chi)}\ ,
\label{eq:ridge_sp_chihat2}
\\
\hq
&=
-\frac{q_0+1}{2r(1+\chi)^2}\ ,
\label{eq:ridge_sp_qhat2}
\\
q_0
&=
\left\langle
\left(
\frac{\lambda\beta+\sqrt{-2\hq}\,z}
{2\hchi+\lambda}
\right)^2
\right\rangle_{z,\beta}\ ,
\label{eq:ridge_sp_q0_raw}
\\
\chi
&=
\frac{
\left\langle
z
\left(
\frac{\lambda\beta+\sqrt{-2\hq}\,z}
{2\hchi+\lambda}
\right)
\right\rangle_{z,\beta}
}{
\sqrt{-2\hq}
}\ .
\label{eq:ridge_sp_chi_raw}
\end{align}
Since \(\langle z\rangle_z=0\), \(\langle z^2\rangle_z=1\), and \(z\) is
independent of \(\beta\), these simplify to
\begin{align}
\hchi
&=
\frac{1}{2r(1+\chi)}\ ,
\label{eq:ridge_sp_chihat}
\\
\hq
&=
-\frac{q_0+1}{2r(1+\chi)^2}\ ,
\label{eq:ridge_sp_qhat}
\\
q_0
&=
\frac{
\lambda^2\overline{\beta^2}
-
2\hq
}{
(2\hchi+\lambda)^2}
\ ,
\label{eq:ridge_q0_sp}
\\
\chi
&=
\frac{1}{2\hchi+\lambda}\ .
\label{eq:ridge_chi_sp}
\end{align}
Here
\begin{equation}
\overline{\beta^2}
=
\int \dd\beta\,p(\beta)\beta^2.
\end{equation}

The first and fourth equations can then be solved simultaneously to yield $\chi=\varphi_1(r,\lambda)$ and $\hat\chi=\varphi_2(r,\lambda)$, with
\begin{align}
    \varphi_1(r,\lambda) &=\frac{\sqrt{4 \lambda  r^2+((\lambda -1) r+1)^2}-\lambda  r+r-1}{2 \lambda  r}\\
    \varphi_2(r,\lambda) &=\frac{\sqrt{(\lambda +1)^2 r^2+2 (\lambda -1) r+1}-r(\lambda +1) +1}{4 r}\ ,\label{phi1phi2}
\end{align}
where we have selected the solution with $\chi>0$. The remaining two equations can be simplified as
\begin{align}
    -\hat q_0-2 r(q_0+1)  \varphi_2(r,\lambda)^2 &=0\\
    \overline{\beta ^2} \lambda ^2 \varphi_1(r,\lambda)^2-2 \hat q_0 \varphi_1(r,\lambda)^2-q_0 &=0\ ,
\end{align}
which can be solved together to give
\begin{align}
    q_0 &= -\frac{\overline{\beta ^2} \lambda ^2 \varphi_1(r,\lambda)^2+4 r \varphi_1(r,\lambda)^2 \varphi_2(r,\lambda)^2}{4 r \varphi_1(r,\lambda)^2 \varphi_2(r,\lambda)^2-1}\ ,\\
    \hat q_0 &= \frac{2 \left(\overline{\beta ^2} \lambda ^2 r \varphi_1(r,\lambda)^2 \varphi_2(r,\lambda)^2+r \varphi_2(r,\lambda)^2\right)}{4 r \varphi_1(r,\lambda)^2 \varphi_2(r,\lambda)^2-1}\ .
\end{align}
Inserting the saddle-point solutions into the zero-temperature free energy  \eqref{eq:f_general_regulariser} and after simplifications, we eventually get 
\begin{align}
    \lim_{P\to\infty}\frac{1}{P}\langle\mathcal{E}\rangle=\varphi_2(r,\lambda)-\frac{1}{2} \overline{\beta ^2} \lambda  (\lambda  \varphi_1(r,\lambda)-1)\ .\label{FINALeqAverageRSS}
\end{align}
For $\lambda\to 0$, this equation reduces to $\varphi_2(r,0)=(1-r)/(2r)$ for $0<r<1$, in agreement with \eqref{minfstarnolambda}.  In Fig. 2 of the main text (bottom right panel) we show full agreement between these formulae and numerical simulations.  

\subsection{Lasso regularizer}

For the Lasso regularization case we consider 
\begin{equation}
g_\lambda(w)
=
\frac{\lambda}{2}|w|\ .
\label{eq:lasso_regulariser}
\end{equation}
Therefore the single-site potential becomes
\begin{equation}
V_{\mathrm{Lasso}}(w;\beta,z)
=
\hchi(\beta-w)^2
-
\sqrt{-2\hq}\,z(\beta-w)
+
\frac{\lambda}{2}|w|\ .
\label{eq:lasso_potential}
\end{equation}
To minimize this potential, we complete the square in \(w\). Define the
effective scalar field
\begin{equation}
a(\beta,z)
=
\beta
-
\frac{\sqrt{-2\hq}}{2\hchi}\,z\ .
\label{eq:lasso_effective_field}
\end{equation}
Then, up to terms independent of \(w\),
\begin{equation}
V_{\mathrm{Lasso}}(w;\beta,z)
=
\hchi\left(w-a(\beta,z)\right)^2
+
\frac{\lambda}{2}|w|
+
\mathrm{constant}\ .
\label{eq:lasso_square_completed}
\end{equation}
Therefore the minimizer is the soft-thresholding function
\begin{equation}
{
w^*_{\mathrm{Lasso}}(\beta,z)
=
\operatorname{ST}
\left(
a(\beta,z),
\frac{\lambda}{4\hchi}
\right)\ .
}
\label{eq:lasso_wstar}
\end{equation}
Here
\begin{equation}
\operatorname{ST}(a,\theta)
=
\operatorname{sign}(a)(|a|-\theta)_+,
\qquad
(x)_+=\max(x,0)\ .
\label{eq:soft_threshold_definition}
\end{equation}
Equivalently,
\begin{equation}
w^*_{\mathrm{Lasso}}(\beta,z)
=
\begin{cases}
a(\beta,z)-\dfrac{\lambda}{4\hchi},
& a(\beta,z)>\dfrac{\lambda}{4\hchi}\ ,
\\[8pt]
0,
& |a(\beta,z)|\leq\dfrac{\lambda}{4\hchi}\ ,
\\[8pt]
a(\beta,z)+\dfrac{\lambda}{4\hchi},
& a(\beta,z)<-\dfrac{\lambda}{4\hchi}\ .
\end{cases}
\label{eq:lasso_wstar_piecewise}
\end{equation}

The corresponding estimation error entering the overlap is
\begin{equation}
{
\beta-w^*_{\mathrm{Lasso}}(\beta,z)
=
\beta
-
\operatorname{ST}
\left(
\beta
-
\frac{\sqrt{-2\hq}}{2\hchi}\,z,
\frac{\lambda}{4\hchi}
\right)\ .
}
\label{eq:lasso_error_star}
\end{equation}

Substituting \eqref{eq:lasso_error_star} into the general saddle-point
equations gives
\begin{align}
\hchi
&=
\frac{1}{2r(1+\chi)}\ ,
\label{eq:lasso_sp_chihat}
\\
\hq
&=
-\frac{q_0+1}{2r(1+\chi)^2}\ ,
\label{eq:lasso_sp_qhat}
\\
q_0
&=
\left\langle
\left[
\beta
-
\operatorname{ST}
\left(
\beta
-
\frac{\sqrt{-2\hq}}{2\hchi}\,z,
\frac{\lambda}{4\hchi}
\right)
\right]^2
\right\rangle_{z,\beta}\ ,
\label{eq:lasso_sp_q0}
\\
\chi
&=
\frac{
\left\langle
z
\left[
\beta
-
\operatorname{ST}
\left(
\beta
-
\frac{\sqrt{-2\hq}}{2\hchi}\,z,
\frac{\lambda}{4\hchi}
\right)
\right]
\right\rangle_{z,\beta}
}{
\sqrt{-2\hq}
}\ .
\label{eq:lasso_sp_chi}
\end{align}
These equations can be rewritten in equivalent forms involving Error functions: in any case, they need to be solved numerically for any given value of $\lambda$. In Fig. 2 of the main text (bottom right panel) we show full agreement between these formulae and numerical simulations.

\subsection{Order parameter comparison}

The order parameter $q_0$ has a direct interpretation as the error in recovering the teacher coefficients. To see this, recall that the replica overlap was introduced in Eq.~\eqref{eq:defQabinline} as
\begin{equation}
Q_{ab}
=
\frac{1}{P}\sum_{j=1}^{P}
(\beta_j-w_{ja})(\beta_j-w_{jb})\ ,
\end{equation}
and, under the replica-symmetric ansatz of Eq.~\eqref{QabRSassumptiondef},
$Q_{aa}=q_0+\chi$ and $Q_{a\neq b}=q_0$. In the zero-temperature limit
$\gamma\to\infty$, the scaling $\chi\to\chi/\gamma$ implies that the diagonal
and off-diagonal overlaps coincide: all replicas concentrate on the same
minimizer $\bm w^\star$. Consequently,
\begin{equation}
q_0
=
\frac{1}{P}
\left\langle
\left\|\bm{\beta}-\bm w^\star\right\|^2
\right\rangle\ ,
\end{equation}
which is also obtained directly from the saddle-point equation~\eqref{eq:sp_general_chihat}. Thus
$q_0$ is the mean-squared parameter-estimation error per coefficient: it
measures how accurately the fitted model recovers the signal that generated
the data, rather than how accurately it fits the observed responses.

This distinction becomes particularly consequential at the interpolation
threshold. For OLS, Eq.~\eqref{eq:q0OLS} gives
\begin{equation}
q_0=\frac{r}{1-r}\ ,
\end{equation}
which diverges as $r\to1^-$. The origin of this divergence is transparent
from the OLS estimator (see Eq. \eqref{eq:ols_wstar_exact} below). Writing $\bm y=A\bm{\beta}+\bm{\epsilon}$,
\begin{equation}
\bm w^\star-\bm{\beta}
=
(A^\top A)^{-1}A^\top\bm{\epsilon}\ ,
\end{equation}
so fluctuations of the noise are amplified by the inverse Gram matrix.
As $r$ approaches one from below, increasingly small eigenvalues of
$A^\top A$ make the inferred coefficients extremely sensitive to the
particular realization of the noise. At the same point, however, the
training loss decreases to zero [Eq.~\eqref{minfstarnolambda}]. Figure~1 therefore exposes a
sharp separation between fitting and inference: at interpolation, OLS fits
the observed data perfectly while its estimate of the underlying
coefficients becomes singular.

\begin{figure}[H]
\centering
\includegraphics[width=\columnwidth]{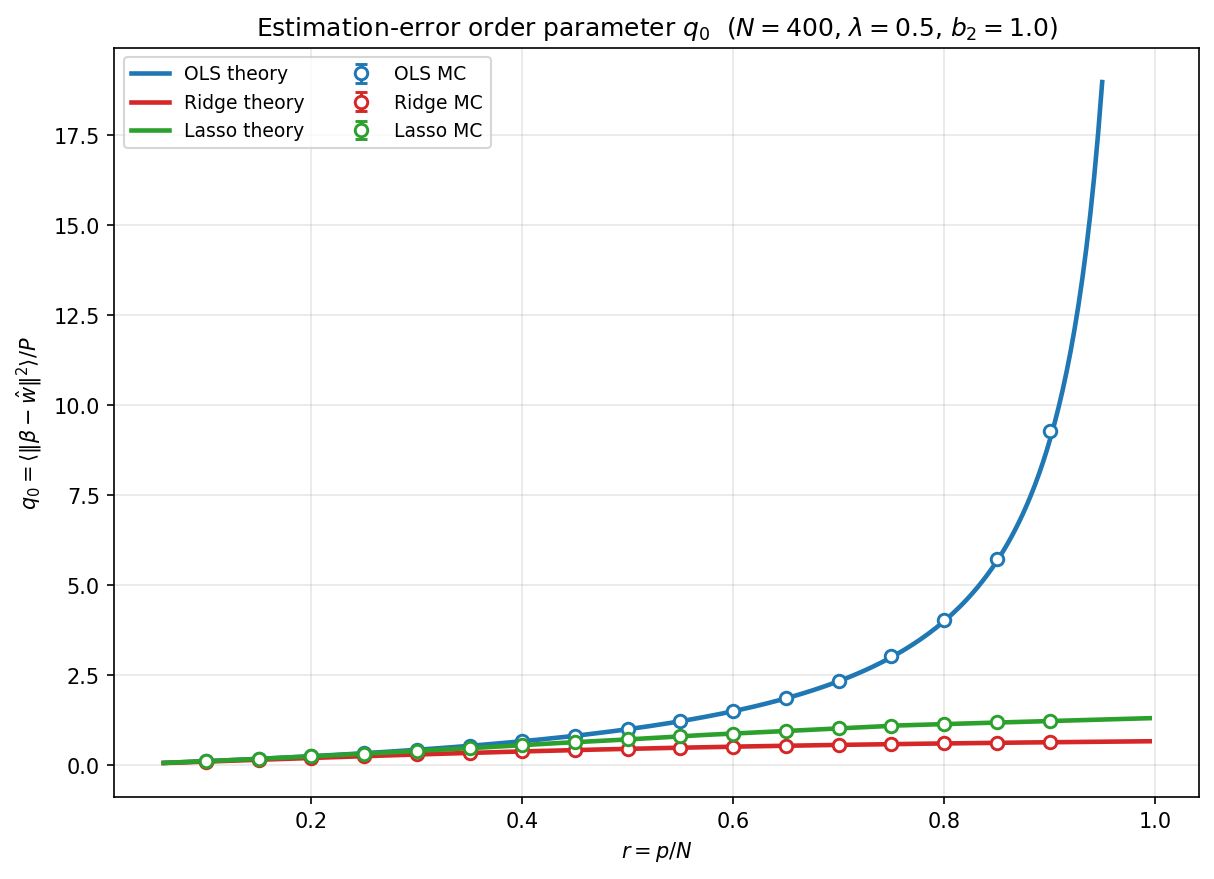}
\caption{Estimation-error order parameter
$q_0$ vs.\ $r=P/N$, for
OLS, Ridge, and Lasso ($N=400$, $\lambda=0.5$ for both regularized
cases, $\bar\beta^2=1$). Solid lines: replica prediction; open circles:
Monte Carlo estimates  from $N_{\mathrm{tot}} = 1.1\times 10^{9}$ (OLS),
$1.1\times 10^{6}$ (Ridge), and $8.9\times 10^{5}$ (Lasso) independent
realizations of the disorder ($X$, $\bm{\epsilon}$, $\bm{\beta}$).}
\label{fig:q0}
\end{figure}

Ridge and Lasso regularizations remove this instability. As shown in
Fig.~\ref{fig:q0}, their corresponding values of $q_0$ remain finite through $r=1$:
penalizing the coefficients suppresses the amplification of poorly
constrained directions and prevents the divergence of the estimation error.
Regularization therefore changes the interpolation threshold from a
singularity in parameter recovery into a smooth crossover, stabilizing the
inferred model even where unregularized least squares becomes maximally
sensitive to the data realization. Figure~\ref{fig:q0} shows $q_0$ for OLS, Ridge, and Lasso
as a function of $r=P/N$, at fixed $\lambda=0.5$ for the regularized cases
and $\overline{\beta^2}=1$. The numerical
simulations (open circles) are in excellent agreement with the replica
predictions (solid lines) for all three cases.

\section{Replica Calculation for the full distribution of the minimal loss}

We now turn from the average minimal loss to its full distribution. The key idea is to set $n=\frac{s}{\gamma}$ before taking the zero-temperature limit \(\gamma\to\infty\). Since
\[
Z(\gamma)
=
\int \dd\bm w\,{\re}^{-\gamma H_\lambda(\bm w)}
\sim
{\re}^{-\gamma \min_{\bm w}H_\lambda(\bm w)} \ ,
\]
we have
\begin{equation}
Z(\gamma)^{s/\gamma}
\sim
{\re}^{-s\min_{\bm w}H_\lambda(\bm w)}
\end{equation}
for large $\gamma$.
Therefore
\begin{equation}
\left\langle Z(\gamma)^{s/\gamma}\right\rangle
\longrightarrow
\left\langle
{\re}^{-s\mathcal{E}}
\right\rangle\ ,\label{ZsgammaLaplacesE}
\end{equation}
which is the cumulant generating function of the minimal loss, i.e., the Laplace transform of its distribution.

Starting from the replica-symmetric expression before expanding only to
linear order in \(n\), we must keep the \(n^2\) terms in
\(\sum_{ab}Q_{ab}\hat Q_{ab}\) (see Eq.\eqref{eq:Pterm}), because after the zero-temperature rescaling this object
contributes at order one when \(n=s/\gamma\). We again use the same zero-temperature rescaling as before:
\begin{equation}
\chi\to\frac{\chi}{\gamma}\ ,
\qquad
\hat q_0\to\gamma^2\hat q_0\ ,
\qquad
\hat\chi\to\gamma\hat\chi\ .
\end{equation}

The determinant contribution (see Eq. \eqref{eq:detG_expansion_general}) becomes
\begin{align}
\log\det G
&=
(n-1)\log(1+\chi)
+
\log\left(1+\chi+n\gamma(q_0+1)\right)
\nonumber\\
&\xrightarrow[n=s/\gamma]{\gamma\to\infty}
-\log(1+\chi)
+
\log\left(1+\chi+s(q_0+1)\right)\ .
\end{align}
Hence
\begin{equation}
-\frac{N}{2}\log\det G
=
-\frac{P}{2r}
\log
\left(
\frac{1+\chi+s(q_0+1)}{1+\chi}
\right)\ .
\end{equation}

The overlap-conjugate term gives
\begin{align}
P\sum_{a,b}\hat Q_{ab}Q_{ab}
&=
P\left[
n(\hat q_0\chi+q_0\hat\chi+\chi\hat\chi)
+
n^2q_0\hat q_0
\right]
\nonumber\\
&\longrightarrow
P\left[
s(\hat q_0\chi+q_0\hat\chi)
+
s^2q_0\hat q_0
\right],
\end{align}
where the term \(s\chi\hat\chi/\gamma\) has vanished.

Finally, the single-site term becomes
\begin{equation}
\left[
\int \mathrm{d}\beta\,p(\beta)
\left\langle
\left(
\int \dd w\,{\re}^{-\gamma V(w;\beta,z)}
\right)^{s/\gamma}
\right\rangle_z
\right]^P,
\end{equation}
where
\begin{equation}
V(w;\beta,z)
=
\hat\chi(\beta-w)^2
-
\sqrt{-2\hat q_0}\,z(\beta-w)
+
g_\lambda(w)\ .
\end{equation}
By Laplace's method,
\begin{equation}
\left(
\int \dd w\,{\re}^{-\gamma V(w;\beta,z)}
\right)^{s/\gamma}
\longrightarrow
{\re}^{-s\,m(\beta,z)}\ ,
\end{equation}
with
\begin{equation}
m(\beta,z)
=
\min_w V(w;\beta,z)\ .
\end{equation}

Therefore
\begin{equation}
\left\langle Z_\gamma^{s/\gamma}\right\rangle
\sim
\int \dd q_0\,\dd\chi\,\dd\hat q_0\,\dd\hat\chi\,
\exp\left[
P\,\mathcal{S}_s(q_0,\chi,\hat q_0,\hat\chi)
\right],\label{ZPSs}
\end{equation}
where the finite-\(s\) action is
\begin{equation}
{
\begin{aligned}
\mathcal{S}_{s}(q_0,\chi,\hat q_0,\hat\chi)
&=
s(\hat q_0\chi+q_0\hat\chi)
+
s^2q_0\hat q_0
-\frac{1}{2r}
\log
\left(
\frac{1+\chi+s(q_0+1)}{1+\chi}
\right)
+
\log
\left\langle
{\re}^{-s\,m(\beta,z)}
\right\rangle_{z,\beta}\ . \\
\end{aligned}
}
\label{sm:eq:Psi_s_general}
\end{equation}
Here
\begin{equation}
\left\langle \cdots \right\rangle_{z,\beta}
=
\int \mathrm{D}z\int \dd\beta\,p(\beta)(\cdots),
\qquad
\mathrm{D}z=
\frac{\dd z}{\sqrt{2\pi}}{\re}^{-z^2/2}\ .
\end{equation}
The function \(\mathcal{S}_s\) is not yet the Laplace-space rate function. It is the
finite-\(s\) replica action before extremization over the order parameters. Taking a saddle-point approach for large $P$ in Eq. \eqref{ZPSs} and appealing to Eq. \eqref{ZsgammaLaplacesE}
\[
\left\langle \re^{-s\mathcal E}\right\rangle
\sim
\re^{-P\Phi(s)},
\qquad
\mathcal E=\min_{\bm w}H_\lambda(\bm w)\ .
\]
Since
\[
\left\langle Z_\gamma^{s/\gamma}\right\rangle
\sim
\int \dd q_0\,\dd\chi\,\dd\hat q_0\,\dd\hat\chi\,
{\re}^{P\mathcal{S}_s(q_0,\chi,\hat q_0,\hat\chi)}\approx {\re}^{P \operatorname*{extr}_{q_0,\chi,\hat q_0,\hat\chi}
\mathcal{S}_{s}(q_0,\chi,\hat q_0,\hat\chi)}\ ,
\]

it follows that
\begin{equation}
{
\Phi(s)
=
-\operatorname*{extr}_{q_0,\chi,\hat q_0,\hat\chi}
\mathcal{S}_{s}(q_0,\chi,\hat q_0,\hat\chi)\ .
}
\label{eq:Phi_s_general}
\end{equation}

\subsection{General finite-\texorpdfstring{$s$}{s} saddle point equations}

The finite-\(s\) action is
\begin{equation}
\mathcal S_s
=
s(\hat q_0\chi+q_0\hat\chi)
+s^2q_0\hat q_0
-\frac{1}{2r}
\log
\frac{1+\chi+s(q_0+1)}{1+\chi}
+
\log
\left\langle {\re}^{-s m(\beta,z)}\right\rangle_{z,\beta}\ ,
\end{equation}
where
\begin{equation}
m(\beta,z)
=
\min_w
\left[
\hat\chi(\beta-w)^2
-
\sqrt{-2\hat q_0}\,z(\beta-w)
+
g_\lambda(w)
\right]\ .
\end{equation}
Let
\begin{equation}
w^*(\beta,z)
=
\operatorname*{argmin}_w
\left[
\hat\chi(\beta-w)^2
-
\sqrt{-2\hat q_0}\,z(\beta-w)
+
g_\lambda(w)
\right]\ ,
\end{equation}
and define
\begin{equation}
u^*(\beta,z)=\beta-w^*(\beta,z)\ .
\end{equation}
We also define the tilted expectation
\begin{equation}
\left\langle F\right\rangle_s
=
\frac{
\left\langle F(\beta,z){\re}^{-s m(\beta,z)}\right\rangle_{z,\beta}
}{
\left\langle {\re}^{-s m(\beta,z)}\right\rangle_{z,\beta}
}\ .
\end{equation}

Using the envelope theorem, the derivatives of \(m\) are
\begin{equation}
\frac{\partial m}{\partial \hat\chi}
=
\left(u^*\right)^2,
\qquad
\frac{\partial m}{\partial \hat q_0}
=
\frac{z u^*}{\sqrt{-2\hat q_0}}\ .
\end{equation}
Therefore the saddle point equations \eqref{eq:f_general_regulariser}--\eqref{eq:sp_general_qhat} reduce to
\begin{align}
\hat\chi+s\hat q_0
&=
\frac{1}{2r\left[1+\chi+s(q_0+1)\right]}\ ,
\\
s\hat q_0
&=
\frac{1}{2r}
\left[
\frac{1}{1+\chi+s(q_0+1)}
-
\frac{1}{1+\chi}
\right]\ ,
\\
q_0
&=
\left\langle
\left(u^*\right)^2
\right\rangle_s\ ,
\\
\chi+s q_0
&=
\frac{
\left\langle
z u^*
\right\rangle_s
}{
\sqrt{-2\hat q_0}
}\ .
\end{align}
Equivalently, the saddle point system can be written as
\begin{align}
\hat\chi
&=
\frac{1}{2r(1+\chi)}\ ,
\\
\hat q_0
&=
\frac{1}{s}
\left[
\frac{1}{2r\left[1+\chi+s(q_0+1)\right]}
-
\hat\chi
\right]\ ,
\\
q_0
&=
\left\langle
\left(\beta-w^*(\beta,z)\right)^2
\right\rangle_s\ ,
\\
\chi+s q_0
&=
\frac{
\left\langle
z\left(\beta-w^*(\beta,z)\right)
\right\rangle_s
}{
\sqrt{-2\hat q_0}
}\ .
\end{align}

\subsection{OLS (no regularizer)}

For ordinary least squares, the regularizer term is not present, i.e., $g_\lambda(w)=0$.
The single-site potential is
\begin{equation}
V_{\mathrm{OLS}}(w;\beta,z)
=
\hat\chi(\beta-w)^2
-
\sqrt{-2\hat q_0}\,z(\beta-w)\ .
\end{equation}
Setting
\begin{equation}
u=\beta-w\ ,
\end{equation}
we have
\begin{equation}
V_{\mathrm{OLS}}(u;z)
=
\hat\chi u^2
-
\sqrt{-2\hat q_0}\,zu\ .
\end{equation}
The minimizer is
\begin{equation}
u^*
=
\beta-w^*
=
\frac{\sqrt{-2\hat q_0}}{2\hat\chi}z\ .
\end{equation}
The corresponding minimum is
\begin{equation}
m_{\mathrm{OLS}}(z)
=
\min_u
\left[
\hat\chi u^2
-
\sqrt{-2\hat q_0}\,zu
\right]
=
\frac{\hat q_0}{2\hat\chi}z^2\ .\label{mOLSzquadro}
\end{equation}
Thus the finite-\(s\) action becomes
\begin{equation}
{
\begin{aligned}
\mathcal{S}_{s}^{\mathrm{OLS}}
&=
s(\hat q_0\chi+q_0\hat\chi)
+
s^2q_0\hat q_0
-\frac{1}{2r}
\log
\left(
\frac{1+\chi+s(q_0+1)}{1+\chi}
\right)
+
\log
\left\langle
\exp\left[
-s\frac{\hat q_0}{2\hat\chi}z^2
\right]
\right\rangle_z .
\end{aligned}
}
\end{equation}
Since \(z\sim\mathcal N(0,1)\),
\begin{equation}
\left\langle
\exp\left[
-s\frac{\hat q_0}{2\hat\chi}z^2
\right]
\right\rangle_z
=
\left(
1+s\frac{\hat q_0}{\hat\chi}
\right)^{-1/2},\label{OLSsinglesitezaverage}
\end{equation}
whenever the Gaussian integral is well-defined. Hence
\begin{equation}
{
\mathcal{S}_{s}^{\mathrm{OLS}}
=
s(\hat q_0\chi+q_0\hat\chi)
+
s^2q_0\hat q_0
-
\frac{1}{2r}
\log
\left(
\frac{1+\chi+s(q_0+1)}{1+\chi}
\right)
-
\frac12
\log
\left(
1+s\frac{\hat q_0}{\hat\chi}
\right).
}
\label{eq:Psi_s_OLS}
\end{equation}
The scaled cumulant generating function is obtained by extremizing this action:
\begin{equation}
{
\Phi_{\rm OLS}(s)
=
-\operatorname*{extr}_{q_0,\chi,\hat q_0,\hat\chi}
\mathcal{S}_{s}^{\mathrm{OLS}}
(q_0,\chi,\hat q_0,\hat\chi)}\ .\label{eq:Psi_s_OLSv2}
\end{equation}
Let
\begin{equation}
x=1+\chi+s(q_0+1)\ .
\end{equation}
The saddle-point equations are
\begin{align}
\frac{\partial \mathcal S_s^{\rm OLS}}{\partial q_0}=0
&\quad\Rightarrow\quad
s(\hat\chi+s\hat q_0)
-
\frac{s}{2rx}
=0\ ,
\\
\frac{\partial \mathcal S_s^{\rm OLS}}{\partial \chi}=0
&\quad\Rightarrow\quad
s\hat q_0
-
\frac{1}{2r}
\left[
\frac{1}{x}
-
\frac{1}{1+\chi}
\right]
=0\ ,
\\
\frac{\partial \mathcal S_s^{\rm OLS}}{\partial \hat q_0}=0
&\quad\Rightarrow\quad
s\chi+s^2q_0
-
\frac{s}{2\hat\chi}
\frac{1}{1+s\hat q_0/\hat\chi}
=0\ ,
\\
\frac{\partial \mathcal S_s^{\rm OLS}}{\partial \hat\chi}=0
&\quad\Rightarrow\quad
sq_0
+
\frac{s\hat q_0}{2\hat\chi^2}
\frac{1}{1+s\hat q_0/\hat\chi}
=0\ .
\end{align}

Equivalently,
\begin{align}
\hat\chi+s\hat q_0
&=
\frac{1}{2rx}\ ,
\\
s\hat q_0
&=
\frac{1}{2r}
\left[
\frac{1}{x}
-
\frac{1}{1+\chi}
\right]\ ,
\\
\chi+sq_0
&=
\frac{1}{2\hat\chi}
\frac{1}{1+s\hat q_0/\hat\chi}\ ,
\\
q_0
&=
-\frac{\hat q_0}{2\hat\chi^2}
\frac{1}{1+s\hat q_0/\hat\chi}\ .
\end{align}
The first two equations imply
\begin{equation}
{
\hat\chi=\frac{1}{2r(1+\chi)}
}\ .
\end{equation}

\paragraph{Evaluation of \(\Phi_{\rm OLS}(s)\).}

Starting from Eq. \eqref{eq:Psi_s_OLS} we define $x=1+\chi+s(q_0+1).$
From the saddle point equations we have
\begin{equation}
\hat\chi+s\hat q_0=\frac{1}{2rx }\ ,
\end{equation}
and
\begin{equation}
\hat\chi=\frac{1}{2r(1+\chi)} \ .
\end{equation}
Moreover,
\begin{equation}
\chi+s q_0
=
\frac{1}{2(\hat\chi+s\hat q_0)}\ .
\end{equation}
Using \(\hat\chi+s\hat q_0=1/(2rx)\), this gives
\begin{equation}
\chi+s q_0=rx\ .
\end{equation}
Since $
x=
1+s+\chi+s q_0$ we obtain
\begin{equation}
x=1+s+rx\ .
\end{equation}
Therefore,
\begin{equation}
{
x(s)=\frac{1+s}{1-r}\ .
}
\end{equation}

The remaining saddle values are
\begin{equation}
\chi^*=\frac{r}{1-r},
\qquad
\hat\chi^*=\frac{1-r}{2r}\ ,
\end{equation}
and
\begin{equation}
q_0^*=\frac{r}{1-r},
\qquad
\hat q_0^*(s)
=
-\frac{1-r}{2r(1+s)}\ .
\end{equation}

Substituting the saddle solution into \(\mathcal S_s^{\rm OLS}\), one finds
\begin{equation}
\mathcal S_s^{\rm OLS}\big|_{\rm saddle}
=
-\frac{1-r}{2r}\log(1+s)\ .
\end{equation}
Hence, since
\begin{equation}
\Phi_{\rm OLS}(s)
=
-\mathcal S_s^{\rm OLS}\big|_{\rm saddle}\ ,
\end{equation}
we obtain
\begin{equation}
{
\Phi_{\rm OLS}(s)
=
\frac{1-r}{2r}\log(1+s)
}\ .
\label{eq:Phi_OLS_final}
\end{equation}

Differentiating and setting $s=0$, we get
\begin{equation}
 \Phi_{\rm OLS}'(0)=\frac{1-r}{2r}\ ,   
\end{equation}
which coincides with the average value of the OLS loss derived earlier in Eq. \eqref{minfstarnolambda}, as predicted by the general theory of large deviations.

\subsection{Ridge Regularizer}

We now consider the Ridge regularizer
\begin{equation}
g_\lambda(w)
=
\frac{\lambda}{2}w^2\ .
\end{equation}
In this case, the single-site potential reads
\begin{equation}
V_{\mathrm{Ridge}}(w;\beta,z)
=
\hat\chi(\beta-w)^2
-
\sqrt{-2\hat q_0}\,z(\beta-w)
+
\frac{\lambda}{2}w^2\ .
\end{equation}
Equivalently, with \(u=\beta-w\),
\begin{equation}
V_{\mathrm{Ridge}}(u;\beta,z)
=
\hat\chi u^2
-
\sqrt{-2\hat q_0}\,zu
+
\frac{\lambda}{2}(\beta-u)^2\ .
\end{equation}
The minimizer is
\begin{equation}
u^*_{\mathrm{Ridge}}
=
\beta-w^*_{\mathrm{Ridge}}
=
\frac{
\lambda\beta+\sqrt{-2\hat q_0}\,z
}{
2\hat\chi+\lambda
}\ .
\end{equation}
The minimum value is therefore
\begin{equation}
m_{\mathrm{Ridge}}(\beta,z)
=
\frac{\lambda}{2}\beta^2
-
\frac{
\left(
\lambda\beta+\sqrt{-2\hat q_0}\,z
\right)^2
}{
2(2\hat\chi+\lambda)
}\ .
\label{eq:m_Ridge}
\end{equation}
Therefore
\begin{equation}
{
\begin{aligned}
\mathcal{S}_{s}^{\mathrm{Ridge}}
&=
s(\hat q_0\chi+q_0\hat\chi)
+
s^2q_0\hat q_0
-\frac{1}{2r}
\log
\left(
\frac{1+\chi+s(q_0+1)}{1+\chi}
\right)
+
\log
\left\langle
\exp\left[
-s\,m_{\mathrm{Ridge}}(\beta,z)
\right]
\right\rangle_{z,\beta}\ .
\end{aligned}
}
\label{eq:Psi_s_Ridge}
\end{equation}
Explicitly,
\begin{equation}
\left\langle
{\rm e}^{-s\,m_{\mathrm{Ridge}}(\beta,z)}
\right\rangle_{z,\beta}
=
\left\langle
\exp\left[
-\frac{s\lambda}{2}\beta^2
+
\frac{s}{2(2\hat\chi+\lambda)}
\left(
\lambda\beta+\sqrt{-2\hat q_0}\,z
\right)^2
\right]
\right\rangle_{z,\beta}\ .
\end{equation}

We introduce the shorthand
\begin{equation}
A=2\hat\chi+\lambda,
\qquad
B=-2\hat q_0\ .
\end{equation}
Then
\begin{equation}
\lambda\beta+\sqrt{-2\hat q_0}\,z
=
\lambda\beta+\sqrt{B}\,z\ .
\end{equation}
For fixed \(\beta\), the \(z\)-average is Gaussian:
\begin{align}
\left\langle
\exp\left[
\frac{s}{2A}
\left(
\lambda\beta+\sqrt{B}\,z
\right)^2
\right]
\right\rangle_z
=
\frac{1}{\sqrt{1-\frac{sB}{A}}}
\exp\left[
\frac{s\lambda^2}{2(A-sB)}\beta^2
\right]\ .
\end{align}
Therefore
\begin{align}
\left\langle
{\re}^{-s\,m_{\mathrm{Ridge}}(\beta,z)}
\right\rangle_{z,\beta}
&=
\frac{1}{\sqrt{1-\frac{sB}{A}}}
\left\langle
\exp\left[
-\frac{s\lambda}{2}\beta^2
+
\frac{s\lambda^2}{2(A-sB)}\beta^2
\right]
\right\rangle_\beta.
\end{align}
Restoring \(A=2\hat\chi+\lambda\) and \(B=-2\hat q_0\), this becomes
\begin{equation}
{
\left\langle
{\re}^{-s\,m_{\mathrm{Ridge}}(\beta,z)}
\right\rangle_{z,\beta}
=
\frac{1}{
\sqrt{
1+\dfrac{2s\hat q_0}{2\hat\chi+\lambda}
}
}
\left\langle
\exp\left[
\left(
-\frac{s\lambda}{2}
+
\frac{s\lambda^2}{
2(2\hat\chi+\lambda+2s\hat q_0)
}
\right)
\beta^2
\right]
\right\rangle_\beta\ .
}
\label{eq:ridge_laplace_factor_z_done}
\end{equation}

It is convenient to introduce the moment-generating function of \(\beta^2\),
\begin{equation}
M_{\beta^2}(u)
=
\left\langle {\re}^{u\beta^2}\right\rangle_\beta
=
\int \mathrm{d}\beta\,p(\beta){\re}^{u\beta^2}\ .
\end{equation}

Then
\begin{equation}
{
\left\langle
{\re}^{-s\,m_{\mathrm{Ridge}}(\beta,z)}
\right\rangle_{z,\beta}
=
\frac{1}{
\sqrt{
1+\dfrac{2s\hat q_0}{2\hat\chi+\lambda}
}
}
M_{\beta^2}
\left(C_s\right)\ ,
}
\label{eq:ridge_laplace_factor_compact}
\end{equation}
with
\begin{equation}
C_s
=
-\frac{s\lambda}{2}
+
\frac{s\lambda^2}{2(\lambda+2\hat\chi+2s\hat q_0)}\ .
\label{eq:Cs_definition}
\end{equation}

The finiteness of $M_{\beta^2}(u)$ is an important restriction on the domain of the
large-deviation calculation. In particular, Eq.~\eqref{eq:ridge_laplace_factor_compact} is well defined only when
$M_{\beta^2}(C_s)<\infty$. On the physical saddle,
$C_s>0$ on the $s<0$ branch associated with the upper-loss tail, so that this
requires a positive exponential moment of $\beta^2$. If $p(\beta)$ is too
heavy-tailed for such a moment to exist, the conventional $P$-speed
large-deviation description cannot in general be continued into this regime.
Rare fluctuations may instead be dominated by one or a few exceptionally large
teacher coefficients, leading to a single-big-jump or condensation mechanism (see main text for a short discussion). The corresponding probability may then decay
with a slower, tail-dependent speed rather than as $\exp[-P\psi(\xi)]$. Thus, for Ridge regression, the analytic domain of
$M_{\beta^2}$ directly determines the range over which the standard
large-deviation principle with speed $P$ can be confidently invoked.

Thus the Ridge finite-\(s\) replica action can be written as
\begin{align}
\mathcal{S}_{s}^{\mathrm{Ridge}}
&=
s(\hat q_0\chi+q_0\hat\chi)
+
s^2q_0\hat q_0
-
\frac{1}{2r}
\log
\left(
\frac{1+\chi+s(q_0+1)}{1+\chi}
\right)
\nonumber\\
&\quad
-\frac{1}{2}
\log
\left(
1+\frac{2s\hat q_0}{2\hat\chi+\lambda}
\right)
+
\log
M_{\beta^2}
\left(
-\frac{s\lambda}{2}
+
\frac{s\lambda^2}{
2(2\hat\chi+\lambda+2s\hat q_0)
}
\right)\ .
\label{eq:Psi_s_Ridge_compact}
\end{align}

The Laplace-space rate function is obtained by extremizing this action:
\begin{equation}
\Phi_{\mathrm{Ridge}}(s)
=
-\operatorname*{extr}_{q_0,\chi,\hat q_0,\hat\chi}
\mathcal{S}_{s}^{\mathrm{Ridge}}\ .
\end{equation}

After performing the Gaussian average over \(z\), but keeping the annealed
average over \(\beta\), the scaled cumulant generating function reads
\begin{align}
\Phi_{\mathrm{Ridge}}(s)
&=
-s(\hat q_0\chi+q_0\hat\chi)
-s^2q_0\hat q_0
\nonumber\\
&\quad
-\frac{1}{2r}\log(1+\chi)
+\frac{1}{2r}\log\left(1+\chi+s(q_0+1)\right)
\nonumber\\
&\quad
-\frac12\log(\lambda+2\hat\chi)
+\frac12\log(\lambda+2\hat\chi+2s\hat q_0)
\nonumber\\
&\quad
-
\log
\int \dd\beta\,p(\beta)
\exp\left[
C_s\,\beta^2
\right]\ .
\label{eq:S_ann_Ridge}
\end{align}

To obtain the saddle point equations we introduce:
\begin{equation}
x
=
1+\chi+s(q_0+1),
\qquad
y
=
\lambda+2\hat\chi+2s\hat q_0\ .
\label{eq:xy_ann}
\end{equation}
Also define the tilted second moment
\begin{equation}
B(C)
=
\frac{\dd}{\dd C}\log M_{\beta^2}(C)
=
\frac{
\int \dd\beta\,p(\beta)\,\beta^2 \mathrm{e}^{C\beta^2}
}{
\int \dd\beta\,p(\beta)\,\mathrm{e}^{C\beta^2}
}\ .
\label{eq:tilted_beta_second_moment}
\end{equation}
Then the saddle point equations are
\begin{align}
&-\hat\chi-s\hat q_0+\frac{1}{2rx}
=0\ ,
\label{eq:sp_ann_q0}
\\
&-s\hat q_0-\frac{1}{2r(1+\chi)}+\frac{1}{2rx}
=0\ ,
\label{eq:sp_ann_chi}
\\
&-\chi-sq_0+\frac{1}{y}
+
s\lambda^2\frac{B(C_s)}{y^2}
=0\ ,
\label{eq:sp_ann_qhat}
\\
&-sq_0-\frac{1}{\lambda+2\hat\chi}
+\frac{1}{y}
+
s\lambda^2\frac{B(C_s)}{y^2}
=0\ .
\label{eq:sp_ann_chihat}
\end{align}

Subtracting the second equation from the first gives
\begin{equation}
{
\hat\chi
=
\frac{1}{2r(1+\chi)}\ .
}
\end{equation}
Subtracting the fourth equation from the third gives
\begin{equation}
{
\chi
=
\frac{1}{\lambda+2\hat\chi}\ .
}
\end{equation}
Therefore, \(\chi\) and \(\hat\chi\) are again independent of \(s\), and coincide
with the average-loss values. Accordingly, we write again
\begin{equation}
\chi=\varphi_1(r,\lambda)\ ,
\qquad
\hat\chi=\varphi_2(r,\lambda)\ ,
\end{equation}
where
\begin{equation}
\varphi_2
=
\frac{1}{2r(1+\varphi_1)},
\qquad
\varphi_1
=
\frac{1}{\lambda+2\varphi_2}\ .
\end{equation}

The remaining equations can be reduced to a single scalar equation for \(x\equiv x(s)\). From
\eqref{eq:sp_ann_q0},
\begin{equation}
\hat\chi+s\hat q_0
=
\frac{1}{2rx}\ ,
\end{equation}
and therefore
\begin{equation}
y(s)
=
\lambda+\frac{1}{r x(s)}\ .
\label{eq:y_ann_x}
\end{equation}
From \eqref{eq:sp_ann_qhat},
\begin{equation}
\chi+sq_0
=
\frac{1}{y}
+
s\lambda^2\frac{B(C_s)}{y^2}\ .
\end{equation}
Using $
x=1+\chi+s(q_0+1)$
we obtain
\begin{equation}
{
x(s)
=
1+s
+
\frac{1}{y(s)}
+
s\lambda^2\frac{B(C_s)}{y(s)^2}\ .
}
\label{eq:x_ann_y}
\end{equation}
Together with \eqref{eq:y_ann_x}, this gives the single implicit equation
\begin{equation}
{
x(s)
=
1+s
+
\frac{1}{\lambda+\dfrac{1}{r x(s)}}
+
s\lambda^2
\frac{
B(C_s)
}{
\left(
\lambda+\dfrac{1}{r x(s)}
\right)^2
}\ ,
}
\label{eq:x_ann_final}
\end{equation}
where
\begin{equation}
{
C_s
=
-\frac{s\lambda}{2}
+
\frac{s\lambda^2}{
2\left(
\lambda+\dfrac{1}{r x(s)}
\right)
}\ .
}
\label{eq:Cs_x_ann}
\end{equation}
The correct branch is chosen by imposing
\begin{equation}
x(0)=1+\varphi_1\ .
\end{equation}
The remaining order parameters are
\begin{equation}
q_0(s)
=
\frac{x(s)-1-\varphi_1-s}{s}\ ,
\qquad
\hat q_0(s)
=
\frac{y(s)-\lambda-2\varphi_2}{2s}\ ,
\end{equation}
with the limit \(s\to0\) taken by continuity.

\paragraph{Annealed Laplace-space rate function.}

\begin{figure}
    \centering
    \includegraphics[width=0.7\linewidth]{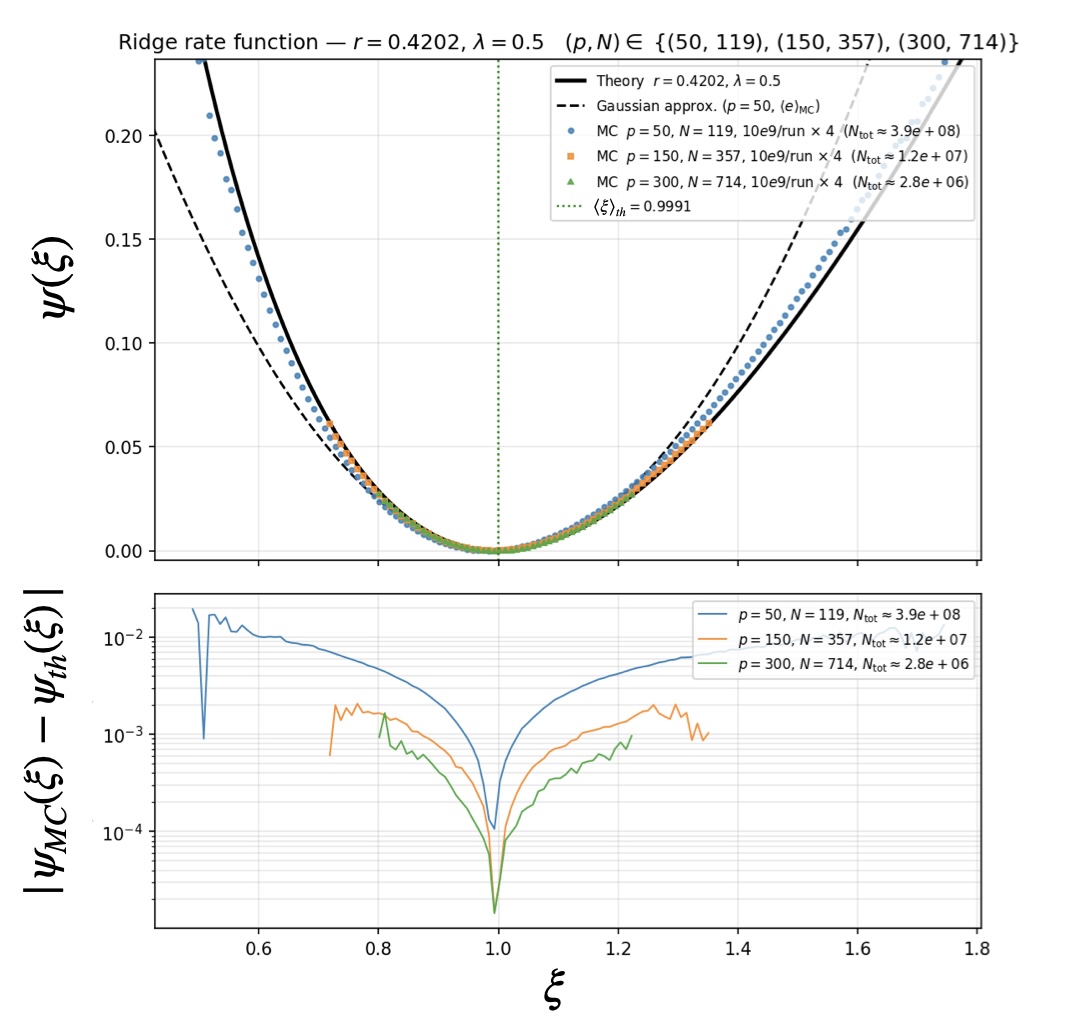}
    \caption{Finite-size collapse of the large-deviation rate function for Ridge regression at fixed ratio $r=P/N=0.4202$ and $\lambda=0.5$. Top: rate
function $\psi(\xi)$ of the intensive loss $\xi=\mathcal E/P$. Solid line:
exact replica prediction; dashed line: Gaussian approximation
about the typical value, constructed from the measured mean at $P=50$;
symbols: rate function estimated from Monte Carlo sampling of the disorder
$(\bm X,\bm\epsilon,\bm\beta)$ at three system sizes along the same ray in
the $(P,N)$ plane, $(P,N)=(50,119)$, $(150,357)$ and $(300,714)$, with
$N_{\rm tot}\simeq3.9\times10^{8}$, $1.2\times10^{7}$ and
$2.8\times10^{6}$ independent realizations respectively. The dotted vertical
line marks the predicted typical value
$\langle\xi\rangle=\Phi'(0)=0.9991$. Since $\psi$ is the rate of a law
exponential in $P$, $\mathbb P(\mathcal E=P\xi)\asymp \mathrm{e}^{-P\psi(\xi)}$, the
three data sets must collapse onto a single curve, as they do; the accessible
range of $\xi$ shrinks with increasing $P$ because at fixed sampling budget
the tails become exponentially harder to reach. Bottom: absolute deviation
$|\psi_{\rm MC}(\xi)-\psi_{\rm th}(\xi)|$ on a logarithmic scale. As we can see, the
residual decreases with $P$ even though the number of realizations decreases,
indicating that it is dominated by $O(1/P)$ finite-size corrections to the
large-deviation limit rather than by sampling noise.}
    \label{fig:ridgecheck}
\end{figure}

Substituting the saddle-point solution into
\eqref{eq:S_ann_Ridge}, we obtain

\begin{equation}
{
\begin{aligned}
\Phi_{\rm Ridge}(s)
&=
\varphi_1\varphi_2
-
\frac{x(s)-1-s}{2r\,x(s)}
+
\frac{1}{2r}
\log\frac{x(s)}{1+\varphi_1}
\\
&\quad
+
\frac12
\log
\frac{
\lambda+\dfrac{1}{r x(s)}
}{
\lambda+2\varphi_2
}
-
\log
\int \dd\beta\,p(\beta)
\exp\left[
\left(
-\frac{s\lambda}{2}
+
\frac{s\lambda^2}{
2\left(\lambda+\dfrac{1}{r x(s)}\right)
}
\right)
\beta^2
\right].
\end{aligned}
}
\label{eq:Phi_Ridge_ann_x_only}
\end{equation}
By construction, the branch \(x(0)=1+\varphi_1\) gives
\begin{equation}
\Phi_{\rm Ridge}(0)=0
\end{equation}
as should be by normalization. Taking the derivative with respect to $s$ and setting $s=0$, one obtains that $\Phi_{\rm Ridge}'(0)$ coincides with the average minimal loss computed in Eq. \eqref{FINALeqAverageRSS}. A numerical test of the validity of the theory is provided in Fig. \ref{fig:ridgecheck}.

\subsection{Lasso Regularizer}

For Lasso, with the convention used here we have $g_\lambda(w)
=
\frac{\lambda}{2}|w|$. The single-site potential is
\begin{equation}
V_{\mathrm{Lasso}}(w;\beta,z)
=
\hat\chi(\beta-w)^2
-
\sqrt{-2\hat q_0}\,z(\beta-w)
+
\frac{\lambda}{2}|w|\ .
\end{equation}
Define the effective scalar field
\begin{equation}
a(\beta,z)
=
\beta
-
\frac{\sqrt{-2\hat q_0}}{2\hat\chi}z\ .
\end{equation}
Then, up to a term independent of \(w\),
\begin{equation}
V_{\mathrm{Lasso}}(w;\beta,z)
=
\hat\chi(w-a(\beta,z))^2
+
\frac{\lambda}{2}|w|
+
\frac{\hat q_0}{2\hat\chi}z^2\ .
\end{equation}
The minimizer is the soft-thresholding function
\begin{equation}
w^*_{\mathrm{Lasso}}(\beta,z)
=
\operatorname{ST}
\left(
a(\beta,z),
\frac{\lambda}{4\hat\chi}
\right)\ ,
\end{equation}
where
\begin{equation}
\operatorname{ST}(a,\theta)
=
\operatorname{sign}(a)(|a|-\theta)_+\ .
\end{equation}

The minimum value can be written as
\begin{equation}
m_{\mathrm{Lasso}}(\beta,z)
=
\frac{\hat q_0}{2\hat\chi}z^2
+
\mathcal M_{\lambda,\hat\chi}(a(\beta,z))\ ,
\label{eq:m_Lasso}
\end{equation}
where
\begin{equation}
\mathcal M_{\lambda,\hat\chi}(a)
=
\min_w
\left[
\hat\chi(w-a)^2+\frac{\lambda}{2}|w|
\right]\ .
\end{equation}
Equivalently, this scalar minimum is the piecewise function
\begin{equation}
\mathcal M_{\lambda,\hat\chi}(a)
=
\begin{cases}
\hat\chi a^2,
&
|a|\leq \dfrac{\lambda}{4\hat\chi}\ ,
\\[8pt]
\dfrac{\lambda}{2}|a|
-
\dfrac{\lambda^2}{16\hat\chi},
&
|a|> \dfrac{\lambda}{4\hat\chi}\ .
\end{cases}
\label{eq:Moreau_Lasso}
\end{equation}

Therefore the finite-\(s\) action for Lasso is
\begin{equation}
{
\begin{aligned}
\mathcal{S}_{s}^{\mathrm{Lasso}}
&=
s(\hat q_0\chi+q_0\hat\chi)
+
s^2q_0\hat q_0
-\frac{1}{2r}
\log
\left(
\frac{1+\chi+s(q_0+1)}{1+\chi}
\right)
+
\log
\left\langle
\exp\left[
-s\,m_{\mathrm{Lasso}}(\beta,z)
\right]
\right\rangle_{z,\beta}.
\end{aligned}
}
\label{eq:Psi_s_Lasso}
\end{equation}
That is,
\begin{equation}
\left\langle
{\re}^{-s\,m_{\mathrm{Lasso}}(\beta,z)}
\right\rangle_{z,\beta}
=
\left\langle
\exp\left[
-s
\left(
\frac{\hat q_0}{2\hat\chi}z^2
+
\mathcal M_{\lambda,\hat\chi}
\left(
\beta
-
\frac{\sqrt{-2\hat q_0}}{2\hat\chi}z
\right)
\right)
\right]
\right\rangle_{z,\beta}.\label{tiltedsinglesiteLasso}
\end{equation}

The finiteness of the tilted single-site average in Eq.~\eqref{tiltedsinglesiteLasso} again imposes a
condition on the tails of $p(\beta)$. In the Lasso case this condition is
weaker than for Ridge. Indeed, from Eq.~\eqref{eq:Moreau_Lasso},
\begin{equation}
M_{\lambda,\widehat{\chi}}(a)
\sim \frac{\lambda}{2}|a|,
\qquad |a|\to\infty ,
\end{equation}
and, since $a(\beta,z)=\beta-kz$ with $z$ Gaussian, the large-$|\beta|$
behavior of the tilted integral is controlled by
$\exp[-s\lambda|\beta|/2]$. Hence, on the upper-loss branch $s<0$, the
standard $P$-speed large-deviation saddle requires the exponential moment
\begin{equation}
\int \mathrm{d}\beta\,p(\beta)\,
\exp\!\left(\frac{|s|\lambda}{2}|\beta|\right)<\infty
\end{equation}
over the relevant range of $s$. If this condition fails, rare fluctuations
may instead be dominated by one or a few exceptionally large teacher
coefficients, giving rise to a single-big-jump or condensation mechanism, and potentially to a slower,
tail-dependent large-deviation speed. Thus the tail
condition is regularizer-dependent: whereas Ridge probes exponential moments
of $\beta^2$, Lasso probes the weaker exponential integrability of
$|\beta|$.

We can now compute this term more explicitly. Define
\begin{equation}
L=\sqrt{-2\hat q_0},
\qquad
k=\frac{L}{2\hat\chi},
\qquad
\theta=\frac{\lambda}{4\hat\chi},
\qquad
a(\beta,z)=\beta-kz\ .
\end{equation}
Completing the square gives
\begin{equation}
V_{\rm Lasso}(w;\beta,z)
=
\hat\chi(w-a)^2
+
\frac{\lambda}{2}|w|
+
\frac{\hat q_0}{2\hat\chi}z^2\ . 
\end{equation}
Therefore
\begin{equation}
w^*(\beta,z)
=
\operatorname{ST}\left(a(\beta,z),\theta\right),
\qquad
\operatorname{ST}(a,\theta)=\operatorname{sign}(a)(|a|-\theta)_+\ .
\end{equation}
The minimum value is
\begin{equation}
m_{\rm Lasso}(\beta,z)
=
\frac{\hat q_0}{2\hat\chi}z^2
+
\mathcal M_{\lambda,\hat\chi}(a(\beta,z))\ ,
\end{equation}
where
\begin{equation}
\mathcal M_{\lambda,\hat\chi}(a)
=
\begin{cases}
\hat\chi a^2,
&
|a|\leq \theta\ ,
\\[6pt]
\dfrac{\lambda}{2}|a|-\dfrac{\lambda^2}{16\hat\chi},
&
|a|>\theta\ .
\end{cases}
\end{equation}

For fixed \(\beta\), the thresholds in \(z\) are
\begin{equation}
z_-(\beta)=\frac{\beta-\theta}{k},
\qquad
z_+(\beta)=\frac{\beta+\theta}{k}\ .
\end{equation}
The three integration regions are therefore
\begin{equation}
a>\theta \Longleftrightarrow z<z_-(\beta),
\qquad
|a|\leq \theta \Longleftrightarrow z_-(\beta)\leq z\leq z_+(\beta),
\qquad
a<-\theta \Longleftrightarrow z>z_+(\beta)\ .
\end{equation}
Thus
\begin{equation}
\int \mathrm{D}z\,
{\re}^{-s\,m_{\rm Lasso}(\beta,z)}
=
I_-(\beta)+I_0(\beta)+I_+(\beta)\ .
\end{equation}

To write the result compactly, define
\begin{equation}
\mathcal G(A,B,C;\ell,u)
=
\frac{e^{C+B^2/(4A)}}{2\sqrt{2A}}
\left[
\operatorname{erf}\left(\sqrt{A}\left(u-\frac{B}{2A}\right)\right)
-
\operatorname{erf}\left(\sqrt{A}\left(\ell-\frac{B}{2A}\right)\right)
\right]\ ,
\label{eq:G_ABC_def}
\end{equation}
with the convention
\begin{equation}
\operatorname{erf}(+\infty)=1,
\qquad
\operatorname{erf}(-\infty)=-1\ .
\end{equation}
This function is simply
\begin{equation}
\mathcal G(A,B,C;\ell,u)
=
\int_{\ell}^{u}\frac{\dd z}{\sqrt{2\pi}}\,
\exp\left[-Az^2+Bz+C\right].
\end{equation}

The coefficients for the two active regions are
\begin{equation}
A_{\rm a}
=
\frac{1}{2}\left(1+s\frac{\hat q_0}{\hat\chi}\right),
\qquad
B_-
=
\frac{s\lambda k}{2},
\qquad
C_-(\beta)
=
-\frac{s\lambda}{2}\beta
+
\frac{s\lambda^2}{16\hat\chi}\ ,
\end{equation}
and
\begin{equation}
B_+
=
-\frac{s\lambda k}{2},
\qquad
C_+(\beta)
=
\frac{s\lambda}{2}\beta
+
\frac{s\lambda^2}{16\hat\chi}\ .
\end{equation}
The coefficients for the inactive region are
\begin{equation}
A_0=\frac12,
\qquad
B_0(\beta)=s\beta L,
\qquad
C_0(\beta)=-s\hat\chi\beta^2\ .
\end{equation}
Therefore
\begin{align}
I_-(\beta)
&=
\mathcal G\left(
A_{\rm a},
B_-,
C_-(\beta);
-\infty,
z_-(\beta)
\right)\ ,
\\
I_0(\beta)
&=
\mathcal G\left(
A_0,
B_0(\beta),
C_0(\beta);
z_-(\beta),
z_+(\beta)
\right)\ ,
\\
I_+(\beta)
&=
\mathcal G\left(
A_{\rm a},
B_+,
C_+(\beta);
z_+(\beta),
+\infty
\right)\ .
\end{align}

The Lasso scaled cumulant generating function is therefore
\begin{equation}
{
\begin{aligned}
\Phi_{\rm Lasso}(s)
&=
-s(\hat q_0\chi+q_0\hat\chi)
-s^2q_0\hat q_0
+
\frac{1}{2r}
\log
\frac{1+\chi+s(q_0+1)}{1+\chi}
-\log
\int \dd\beta\,p(\beta)
\left[
I_-(\beta)+I_0(\beta)+I_+(\beta)
\right],
\end{aligned}
}
\label{eq:Phi_lasso_ann_general_pbeta}
\end{equation}
evaluated at the saddle point.

\subsection{Check of the \texorpdfstring{$\lambda\to0$}{lambda to zero} limit for Lasso regularization}

We now check that, at least for
\[
p(\beta)=\delta(\beta-1)\ ,
\]
the expression of the Lasso scaled cumulant generating function reduces to the OLS expression in the limit
\(\lambda\to0\).

For \(p(\beta)=\delta(\beta-1)\), the single site minimizer becomes
\begin{equation}
w^*_{\rm Lasso}(1,z)
=
\operatorname{ST}
\left(
1-\frac{\sqrt{-2\hat q_0}}{2\hat\chi}z,
\frac{\lambda}{4\hat\chi}
\right)\ .
\end{equation}
In the limit \(\lambda\to0\), the threshold
\begin{equation}
\theta=\frac{\lambda}{4\hat\chi}
\end{equation}
vanishes, and therefore
\begin{equation}
\operatorname{ST}(a,0)=a\ .
\end{equation}
Hence
\begin{equation}
w^*_{\rm Lasso}(1,z)
\underset{\lambda\to0}{\longrightarrow}
1-\frac{\sqrt{-2\hat q_0}}{2\hat\chi}z\ .
\end{equation}
Therefore the corresponding estimation error is
\begin{equation}
1-w^*_{\rm Lasso}(1,z)
\underset{\lambda\to0}{\longrightarrow}
\frac{\sqrt{-2\hat q_0}}{2\hat\chi}z\ .
\end{equation}

The Lasso single-site minimum converges as \(\lambda\to0\) to
\begin{equation}
\mathcal M_{\lambda,\hat\chi}(a)
\longrightarrow
\min_w \hat\chi(w-a)^2
=
0\ .
\end{equation}
Thus (see Eq. \eqref{mOLSzquadro})
\begin{equation}
m_{\rm Lasso}(1,z)
\underset{\lambda\to0}{\longrightarrow}
m_{\rm OLS}(z)
=
\frac{\hat q_0}{2\hat\chi}z^2\ .
\end{equation}
It follows that the single-site factor reduces to the OLS one (see Eq. \eqref{OLSsinglesitezaverage})
\begin{align}
\int \mathrm{D}z\,
\exp\left[-s\,m_{\rm Lasso}(1,z)\right]
&\underset{\lambda\to0}{\longrightarrow}
\int \mathrm{D}z\,
\exp\left[
-s\frac{\hat q_0}{2\hat\chi}z^2
\right]=
\left(
1+s\frac{\hat q_0}{\hat\chi}
\right)^{-1/2}\ .
\end{align}
Therefore
\begin{equation}
-\log
\int \mathrm{D}z\,
\exp\left[-s\,m_{\rm Lasso}(1,z)\right]
\underset{\lambda\to0}{\longrightarrow}
\frac12
\log
\left(
1+s\frac{\hat q_0}{\hat\chi}
\right)\ .
\end{equation}

Using the limits above for \(p(\beta)=\delta(\beta-1)\), we obtain
\begin{align}
\Phi_{\rm Lasso}^{\delta}(s)
\underset{\lambda\to0}{\longrightarrow}
\Phi_{\rm OLS}(s)
=
\operatorname*{extr}_{q_0,\chi,\hat q_0,\hat\chi}
\Bigg[
&-s(\hat q_0\chi+q_0\hat\chi)
-s^2q_0\hat q_0
\nonumber\\
&+
\frac{1}{2r}
\log
\frac{1+\chi+s(q_0+1)}{1+\chi}
+
\frac12
\log
\left(
1+s\frac{\hat q_0}{\hat\chi}
\right)
\Bigg]\ ,
\end{align}
where $\Phi_{\rm OLS}(s)$ was computed earlier (see \eqref{eq:Psi_s_OLS} and \eqref{eq:Psi_s_OLSv2}).

The saddle point equations reduce consistently as well. In the limit
\(\lambda\to0\),
\begin{equation}
1-w^*_{\rm Lasso}(1,z)
=
\frac{\sqrt{-2\hat q_0}}{2\hat\chi}z\ .
\end{equation}
Therefore
\begin{equation}
q_0
=
\left\langle
(1-w^*)^2
\right\rangle_s
\longrightarrow
\left\langle
\left(
\frac{\sqrt{-2\hat q_0}}{2\hat\chi}z
\right)^2
\right\rangle_s\ ,
\end{equation}
and
\begin{equation}
\chi+s q_0
=
\frac{
\left\langle
z(1-w^*)
\right\rangle_s
}{
\sqrt{-2\hat q_0}
}
\longrightarrow
\frac{
\left\langle
z
\left(
\frac{\sqrt{-2\hat q_0}}{2\hat\chi}z
\right)
\right\rangle_s
}{
\sqrt{-2\hat q_0}
}\ .
\end{equation}
These are exactly the OLS finite-\(s\) saddle point equations, with tilted Gaussian
measure
\begin{equation}
\dd\mu_s(z)
=
\frac{
\mathrm{D}z\,
\exp\left[
-s\frac{\hat q_0}{2\hat\chi}z^2
\right]
}{
\int \mathrm{D}z'\,
\exp\left[
-s\frac{\hat q_0}{2\hat\chi}{z'}^2
\right]
}\  .
\end{equation}

\begin{figure}
    \centering
    \includegraphics[width=0.7\linewidth]{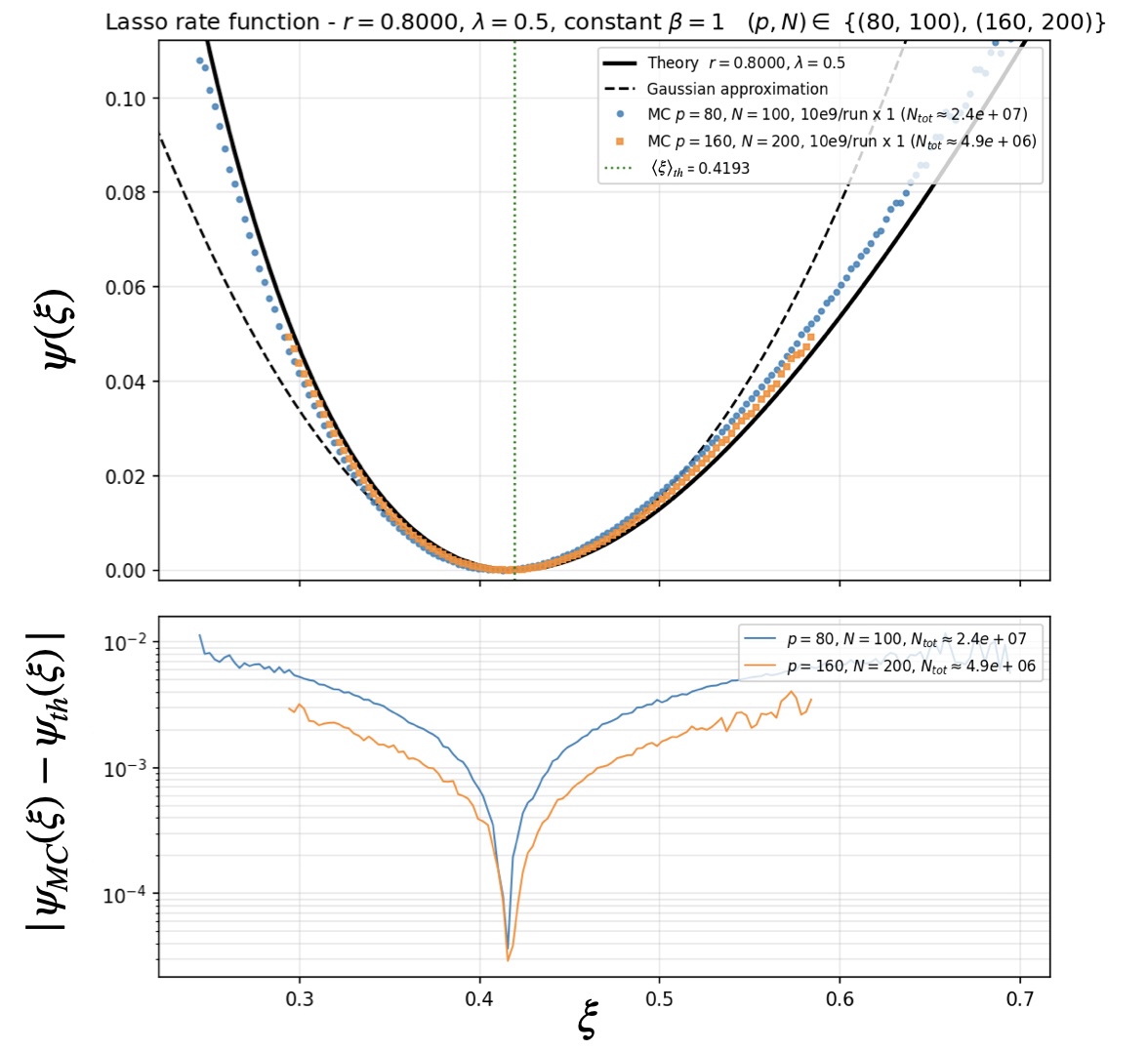}
    \caption{Lasso regression at $r=P/N=0.8000$,
$\lambda=0.5$, with deterministic teacher weights $\beta_j=1$. Top: rate
function $\psi(\xi)$ at $(P,N)=(80,100)$ and $(160,200)$, from
$N_{\rm tot}\simeq2.4\times10^{7}$ and $4.9\times10^{6}$ independent
realizations of the disorder; the dotted line marks
$\langle\xi\rangle=\Phi'(0)=0.4193$. The two sizes collapse onto the replica
prediction over the whole sampled range, confirming that the exponential
scaling in $P$ is already attained at these sizes. The Gaussian
approximation (dashed) lies below the exact rate function on the left of the
typical value and above it on the right: it therefore overestimates the
probability of anomalously good fits and underestimates that of anomalously
bad ones, the deviation growing with distance from $\xi^\star$ in both
directions. Bottom: absolute deviation
$|\psi_{\rm MC}(\xi)-\psi_{\rm th}(\xi)|$, logarithmic scale, again limited
by finite-size rather than statistical error.}
    \label{fig:lassocheck}
\end{figure}
A numerical test of the validity of the theory is provided in Fig. \ref{fig:lassocheck}.

%%%%

\newpage
\section{Mapping between theory and numerical simulations}

The purpose of this section is to make explicit the correspondence between the normalization used in the replica calculation and that used in the numerical simulations, so that the fitted coefficients and regularization parameters can be compared consistently.

Let us adopt the normalization used in the replica calculation above, in which the linear predictor is scaled by $1/\sqrt{P}$. We consider $N$ samples and $P$ features, i.e., $X\in\mathbb{R}^{N\times P}$ with entries $x_j^{(i)}$, and $\bm y\in\mathbb{R}^{N}$ with entries $y^{(i)}$. The model reads
\begin{equation}
y^{(i)}=\frac{1}{\sqrt{P}}\sum_{j=1}^P \beta_j x_j^{(i)}+\epsilon^{(i)},\qquad
\widehat y^{(i)}(\bm w)=\frac{1}{\sqrt{P}}\sum_{j=1}^P w_j x_j^{(i)},\qquad i=1,\dots,N\ .
\end{equation}
In matrix form this is $\bm y=\frac{1}{\sqrt{P}}X\bm\beta+\bm\epsilon$ and $\widehat{\bm y}(\bm w)=\frac{1}{\sqrt{P}}X\bm w$.
Let us focus on the Ridge-regularized case. The Ridge objective is of the form
\begin{align}
 J_{\mathrm{th}}(\bm w)
&=\frac12\Big\|\bm y-\frac{1}{\sqrt{P}}X\bm w\Big\|^2+\frac{\lambda_{\mathrm{th}}}{2}\|\bm w\|^2
\\
&=\frac12\sum_{i=1}^{N}\Big(y^{(i)}-\frac{1}{\sqrt{P}}\sum_{j=1}^P w_j x_j^{(i)}\Big)^2+\frac{\lambda_{\mathrm{th}}}{2}\sum_{j=1}^{P}w_j^2\ ,
\end{align}
where $\lambda_{\mathrm{th}}$ is held $O(1)$ as $P,N\to\infty$ with fixed aspect ratio $r=P/N$.

The minimizer $\bm w^\star$ is obtained by setting the gradient to zero:
\begin{equation}
\nabla_{\bm w} J_{\mathrm{th}}(\bm w)
=-\frac{1}{\sqrt{P}}X^\top\!\Big(\bm y-\frac{1}{\sqrt{P}}X\bm w\Big)+\lambda_{\mathrm{th}}\bm w=0\ ,
\end{equation}
which yields 
\begin{equation}
\Big(\frac{1}{P}X^\top X+\lambda_{\mathrm{th}}I_P\Big)\bm w
=\frac{1}{\sqrt{P}}X^\top\bm y\ ,
\end{equation}
and hence the closed-form Ridge solution in our convention
\begin{equation}
\bm w^\star
=\Big(\frac{1}{P}X^\top X+\lambda_{\mathrm{th}}I_P\Big)^{-1}
\frac{1}{\sqrt{P}}X^\top\bm y\ .
\end{equation}
Here $N$ enters implicitly through the ``data''-dependent quantities
\begin{equation}
X^\top X=\sum_{i=1}^{N} \bm x^{(i)}(\bm x^{(i)})^\top\ ,
\qquad
X^\top\bm y=\sum_{i=1}^{N} \bm x^{(i)}y^{(i)}\ ,
\end{equation}
where $\bm x^{(i)}\in\mathbb{R}^{P}$ denotes the $i$-th row of $X$.

In the numerical simulations, instead, we use the unscaled predictor $\widehat{\bm y}(\bm u)=X\bm u$ and solve
\begin{equation}
J_{\mathrm{num}}(\bm u)
=\frac12\|\bm y-X\bm u\|^2
+\frac{\lambda_{\mathrm{num}}}{2}\|\bm u\|^2
=\frac12\sum_{i=1}^{N}\Big(y^{(i)}-\sum_{j=1}^{P}u_j x_j^{(i)}\Big)^2
+\frac{\lambda_{\mathrm{num}}}{2}\sum_{j=1}^{P}u_j^2\ ,
\end{equation}
whose minimizer $\bm u^\star$ satisfies
\begin{equation}
\big(X^\top X+\lambda_{\mathrm{num}}I_P\big)\bm u=X^\top\bm y,
\qquad\text{i.e.}\qquad
\bm u^\star
=\big(X^\top X+\lambda_{\mathrm{num}}I_P\big)^{-1}X^\top\bm y\ .
\end{equation}

These two formulations are exactly equivalent under the change of variables $\bm w=\sqrt{P}\,\bm u$. In fact, we have
\begin{equation}
J_{\mathrm{th}}(\sqrt{P}\,\bm u)
=\frac12\Big\|\bm y-\frac{1}{\sqrt{P}}X(\sqrt{P}\bm u)\Big\|^2
+\frac{\lambda_{\mathrm{th}}}{2}\|\sqrt{P}\bm u\|^2
=\frac12\|\bm y-X\bm u\|^2
+\frac{\lambda_{\mathrm{th}}P}{2}\|\bm u\|^2\ ,
\end{equation}
so that the parameters must be matched via
\begin{equation}
\lambda_{\mathrm{num}}=\lambda_{\mathrm{th}}\,P
\quad\text{and}\quad
\bm w^\star=\sqrt{P}\,\bm u^\star\ .
\end{equation}
With this identification, the Ridge solutions coincide under the rescaling. Multiplying the numerical normal equations
$(X^\top X+\lambda_{\mathrm{num}}I_P)\bm u=X^\top\bm y$ by $1/\sqrt{P}$ and using $\lambda_{\mathrm{num}}=\lambda_{\mathrm{th}}P$ gives
\begin{equation}
\Big(\frac{1}{P}X^\top X+\lambda_{\mathrm{th}}I_P\Big)(\sqrt{P}\bm u)
=\frac{1}{\sqrt{P}}X^\top\bm y\ ,
\end{equation}
which is exactly the theoretical normal equation with $\bm w=\sqrt{P}\bm u$.
Consequently, predictions agree identically,
\begin{equation}
X\bm u^\star=\frac{1}{\sqrt{P}}X\bm w^\star.
\end{equation}

Note that while $N$ does not appear explicitly in the closed-form expressions above, it controls the behavior of
$\frac{1}{P}X^\top X$ through the aspect ratio $r=P/N$, and hence the asymptotic theory expresses the limiting costs
as functions of $r$.
Therefore, when comparing simulations to the theoretical curves as a function of $r=P/N$, one must either
(i) implement the scaled predictor $X\bm w/\sqrt{P}$ and use $\lambda_{\mathrm{th}}$ directly, or
(ii) keep the unscaled predictor $X\bm u$ and rescale the Ridge parameter as $\lambda_{\mathrm{num}}=\lambda_{\mathrm{th}}P$.

For Lasso, the same change of variables $\bm w=\sqrt{P}\,\bm u$ leaves the predictions unchanged, since
\begin{equation}
\frac{1}{\sqrt{P}}X\bm w = X\bm u\ .
\end{equation}
However, the $\ell_1$ penalty rescales differently from the Ridge penalty:
\begin{equation}
\frac{\lambda_{\mathrm{th}}}{2}\|\bm w\|_1
=
\frac{\lambda_{\mathrm{th}}\sqrt{P}}{2}\|\bm u\|_1\ .
\end{equation}
Therefore, if the numerical Lasso objective is written as
\begin{equation}
J_{\mathrm{num}}^{\mathrm{Lasso}}(\bm u)
=
\frac12\|\bm y-X\bm u\|^2
+
\frac{\lambda_{\mathrm{num}}}{2}\|\bm u\|_1\ ,
\end{equation}
the theoretical and numerical regularization parameters are related by
\begin{equation}
\lambda_{\mathrm{num}}
=
\sqrt{P}\,\lambda_{\mathrm{th}}\ .
\end{equation}

\section{Further consistency checks}

\subsection{OLS (no regularizer) case for $P<N$}

We consider the unregularized problem in the underparameterized regime
$P<N$. In this case the minimization reduces to Ordinary Least Squares (OLS),
for which the distribution of the optimal training loss can be obtained
exactly. This provides a useful benchmark for the replica calculation derived
above.

Consider the OLS problem
\begin{equation}
\mathcal{E}_0
=
\min_{\bm w}\frac12\|\bm y-A\bm w\|^2,
\qquad
A=\frac{X}{\sqrt P}\ .
\label{eq:ols_E0}
\end{equation}

For $P<N$, the Gaussian design matrix $A$ has full column rank $(=P)$ almost surely.
The minimizer is therefore the ordinary least-squares estimator
\begin{equation}
\bm w^\star
=
(A^\top A)^{-1}A^\top\bm y\ .
\label{eq:ols_wstar_exact}
\end{equation}

It is useful to express the corresponding residual geometrically. Let
\begin{equation}
\Pi_A
=
A(A^\top A)^{-1}A^\top \ , 
\label{eq:ols_projector}
\end{equation}
denote the orthogonal projector onto the column space of $A$, with $\Pi_A A=A$. The fitted
response is then $\Pi_A\bm y$, and the optimal loss is
\begin{equation}
\mathcal{E}_0
=
\frac12\|(I_N-\Pi_A)\bm y\|^2\ .
\label{eq:ols_E0_projector}
\end{equation}

Under the teacher--student model, the response is generated as
\begin{equation}
\bm y
=
A\bm\beta+\bm\epsilon,
\qquad
\epsilon_i\sim\mathcal N(0,\sigma^2)\ ,
\label{eq:ols_teacher_student}
\end{equation}
where $\bm\beta$ denotes the true coefficients used to generate the data.
Since $\Pi_A A=A$, it follows that $(I_N-\Pi_A)A=0$. Hence, the residual at the optimum depends only on the noise:
\begin{equation}
\mathcal{E}_0
=
\frac12\bm\epsilon^\top
(I_N-\Pi_A)
\bm\epsilon\ .
\label{eq:ols_E0_noise}
\end{equation}

The matrix $I_N-\Pi_A$ is itself an orthogonal projector, with rank $N-P$. Moreover, since $I_N-\Pi_A$ is a symmetric orthogonal projector, it can be
diagonalized by an orthogonal matrix $U$ as
\begin{equation}
I_N-\Pi_A
=
U
\begin{pmatrix}
I_{N-P} & 0 \\
0 & 0_P
\end{pmatrix}
U^\top\ .
\end{equation}

Writing the Gaussian noise as
\begin{equation}
\bm\epsilon=\sigma\bm z,
\qquad
\bm z\sim\mathcal N(0,I_N)\ ,
\end{equation}
the optimal loss becomes
\begin{equation}
\frac{2\mathcal{E}_0}{\sigma^2}
=
\bm z^\top(I_N-\Pi_A)\bm z\ .
\end{equation}
Using the diagonalization above,
\begin{equation}
\frac{2\mathcal{E}_0}{\sigma^2}
=
(U^\top\bm z)^\top
\begin{pmatrix}
I_{N-P} & 0 \\
0 & 0_P
\end{pmatrix}
(U^\top\bm z)\ .
\end{equation}

As the standard multivariate Gaussian distribution is invariant under
orthogonal transformations,
\begin{equation}
U^\top\bm z\sim\mathcal N(0,I_N)\ .
\end{equation}
Denoting its components again by $z_i$, we therefore obtain
\begin{equation}
\frac{2\mathcal{E}_0}{\sigma^2}
=
\sum_{i=1}^{N-P} z_i^2\ ,
\qquad
z_i\overset{\mathrm{i.i.d.}}{\sim}\mathcal N(0,1)\ .
\end{equation}
By definition, the sum of squares of $N-P$ independent standard Gaussian
variables follows a chi-squared distribution with $N-P$ degrees of freedom and the optimal loss is distributed as
\begin{equation}
\frac{2\mathcal{E}_0}{\sigma^2}\sim\chi^2_{N-P}\ .
\end{equation}

Since this conditional distribution depends only on the rank of the projector
and not on the particular realization of $A$, it is also the unconditional
distribution of the optimal OLS loss.

Equivalently, $\mathcal{E}_0$ is Gamma distributed,
\begin{equation}
\mathcal{E}_0
\sim
\Gamma\left(
\frac{N-P}{2},
\,\sigma^2
\right),
\label{eq:ols_gamma}
\end{equation}
where the second argument is the scale parameter. Hence, its Laplace transform is 
\begin{equation}
\left\langle
\mathrm{e}^{-s\mathcal{E}_0}
\right\rangle
=
(1+s\sigma^2)^{-(N-P)/2}\ .
\label{eq:ols_laplace_exact}
\end{equation}

We now take the double-scaling limit $P,N\to\infty$ at fixed
$r=P/N<1$. Using the definition of the scaled cumulant generating function,
we obtain
\begin{equation}
\Phi_0(s)
=
-\lim_{P\to\infty}\frac1P
\log
\left\langle
\mathrm{e}^{-s\mathcal{E}_0}
\right\rangle
=
\frac{1-r}{2r}
\log(1+s\sigma^2)\ .
\label{eq:ols_Phi_exact}
\end{equation}

For the unit-variance noise considered in the main text,
$\sigma^2=1$, this simplifies to
\begin{equation}
\Phi_0(s)
=
\frac{1-r}{2r}\log(1+s)\ .
\label{eq:ols_Phi_unitvariance}
\end{equation}

The cumulants of the optimal loss follow directly from the Gamma
distribution formula. In the double-scaling limit,
\begin{equation}
\lim_{P\to\infty}
\frac{\kappa_n(\mathcal{E}_0)}{P}
=
\frac{1-r}{2r}
(n-1)!\,\sigma^{2n}\ .
\label{eq:ols_cumulants}
\end{equation}
In particular, the mean and variance per parameter are
\begin{equation}
\lim_{P\to\infty}
\frac{\langle \mathcal{E}_0\rangle}{P}
=
\frac{1-r}{2r}\sigma^2,
\qquad
\lim_{P\to\infty}
\frac{\operatorname{Var}(\mathcal{E}_0)}{P}
=
\frac{1-r}{2r}\sigma^4\ .
\label{eq:ols_mean_variance}
\end{equation}

The same result also determines the large-deviation rate function of the
intensive optimal loss
\begin{equation}
\xi=\frac{\mathcal{E}_0}{P}\ .
\label{eq:ols_xi}
\end{equation}
Using the convention
\begin{equation}
\psi_0(\xi)
=
\sup_s
\left\{
\Phi_0(s)-s\xi
\right\}\ ,
\label{eq:ols_legendre}
\end{equation}
the saddle condition gives
\begin{equation}
1+s^\star\sigma^2
=
\frac{(1-r)\sigma^2}{2r\xi}\ .
\label{eq:ols_sstar}
\end{equation}
Substitution into Eq.~\eqref{eq:ols_legendre} yields
\begin{equation}
\psi_0(\xi)
=
\frac{\xi}{\sigma^2}
-
\frac{1-r}{2r}
-
\frac{1-r}{2r}
\log\left[
\frac{2r\xi}{(1-r)\sigma^2}
\right],
\qquad
\xi>0\ .
\label{eq:ols_rate_general_sigma}
\end{equation}

For unit noise variance this reduces to
\begin{equation}
\psi_0(\xi)
=
\xi
-
\frac{1-r}{2r}
-
\frac{1-r}{2r}
\log\left(
\frac{2r\xi}{1-r}
\right),
\qquad
\xi>0\ .
\label{eq:ols_rate_unitvariance}
\end{equation}
The rate function is minimized at
\begin{equation}
\xi_\star
=
\frac{1-r}{2r}\sigma^2,
\label{eq:ols_typical_loss}
\end{equation}
which coincides with the typical optimal training loss per parameter obtained
from Eq.~\eqref{eq:ols_mean_variance}.

We finally verify that the general replica calculation reproduces this exact
OLS result. Setting $g_\lambda(w)=0$, the single-site potential becomes
\begin{equation}
V_0(w)
=
\hat\chi(\beta-w)^2
-
\sqrt{-2\hat q_0}\,z(\beta-w)\ .
\label{eq:ols_replica_potential}
\end{equation}
Introducing the estimation-error variable
\begin{equation}
u=\beta-w,
\label{eq:ols_u_definition}
\end{equation}
the minimization is elementary and gives
\begin{equation}
u^\star
=
\frac{\sqrt{-2\hat q_0}}{2\hat\chi}\,z,
\qquad
m(\beta,z)
=
\frac{\hat q_0}{2\hat\chi}z^2\ .
\label{eq:ols_replica_minimum}
\end{equation}
Importantly, the minimum is independent of $\beta$. The distribution of the
optimal OLS training loss below the interpolation threshold is therefore
independent of the signal distribution $p(\beta)$, in agreement with the
geometric argument above: after projection onto the orthogonal complement of
the design space, only the noise contributes to the residual.

Substituting Eq.~\eqref{eq:ols_replica_minimum} into the finite-$s$
saddle-point equations gives
\begin{equation}
\chi=q_0=\frac{r}{1-r},
\qquad
\hat\chi=\frac{1-r}{2r},
\qquad
\hat q_0
=
-\frac{\hat\chi}{1+s},
\qquad
r<1\ .
\label{eq:ols_replica_saddle}
\end{equation}

At this saddle, the contributions
$s(\hat q_0\chi+q_0\hat\chi)$ and
$s^2q_0\hat q_0$ cancel. The remaining terms give
\begin{equation}
-\Phi_0(s)
=
-\frac{1}{2r}\log(1+s)
+
\frac12\log(1+s)\ ,
\label{eq:ols_replica_phi_intermediate}
\end{equation}
and hence
\begin{equation}
\Phi_0(s)
=
\frac{1-r}{2r}\log(1+s),
\qquad
r<1\ .
\label{eq:ols_replica_phi_final}
\end{equation}

Equation~\eqref{eq:ols_replica_phi_final} coincides exactly with the asymptotics in Eq.~\eqref{eq:ols_Phi_unitvariance} of the finite-$N$ OLS result. Thus the replica
calculation reproduces not only the typical OLS loss but the full
large-deviation statistics of the optimal training loss in the
underparameterized regime $P<N$. In particular, its typical value is
\begin{equation}
\xi_\star
=
\frac{1-r}{2r},
\qquad
\sigma^2=1\ ,
\label{eq:ols_typical_loss_unitvariance}
\end{equation}
while the complete rate function is given by
Eq.~\eqref{eq:ols_rate_unitvariance}.

\subsection{Replica-free check for Ridge regression with Gaussian teacher weights}
\label{sec:ridge_gaussian_check}

As a further cross-check of the replica calculation, we consider Ridge regression
with Gaussian teacher weights. In this special case the scaled cumulant
generating function of the minimal loss can be obtained without replicas.
The reason is that the Ridge estimator is linear in the response, so the
minimal loss is a quadratic form in the data. For Gaussian teacher weights
and Gaussian noise, the response is itself Gaussian conditional on the
design, and the corresponding Laplace transform can therefore be evaluated
exactly at finite $N$ and $P$.

We retain the normalization used throughout the Supplement and define
\begin{equation}
A=\frac{X}{\sqrt P},
\qquad
K=AA^\top=\frac{XX^\top}{P}\ ,
\end{equation}
so that
\begin{equation}
\bm y=A\bm\beta+\bm\epsilon,
\qquad
\beta_j\sim\mathcal N(0,v),
\qquad
\epsilon^{(i)}\sim\mathcal N(0,\sigma^2)\ ,
\end{equation}
with all variables mutually independent. Here
$v=\overline{\beta^2}$ denotes the variance of the Gaussian teacher prior.
The main derivation corresponds to the case $\sigma^2=1$, but here we consider a generic variance. Since $X$ has i.i.d.\ standard Gaussian entries, $K=XX^\top/P$ is a scaled
Wishart matrix. In the double-scaling limit $N,P\to\infty$ at fixed
$c=N/P=1/r$, its empirical spectral distribution converges to the
Mar\v{c}enko--Pastur law, which will be useful in the following.

For Ridge regression,
\begin{equation}
H_\lambda(\bm w)
=
\frac12\|\bm y-A\bm w\|^2
+
\frac{\lambda}{2}\|\bm w\|^2,
\qquad
\mathcal E
=
\min_{\bm w}H_\lambda(\bm w)\ .
\label{eq:ridge_gaussian_objective}
\end{equation}

Setting the gradient of $H_\lambda$ to zero gives
\begin{equation}
-A^\top(\bm y-A\bm w)+\lambda\bm w=0\ ,
\end{equation}
and hence
\begin{equation}
(A^\top A+\lambda I_P)\hat{\bm w}
=
A^\top\bm y\ .
\label{eq:ridge_gaussian_normal}
\end{equation}
Using the push-through identity
\begin{equation}
(A^\top A+\lambda I_P)^{-1}A^\top
=
A^\top(AA^\top+\lambda I_N)^{-1}\ ,
\end{equation}
the Ridge estimator can equivalently be written as
\begin{equation}
\hat{\bm w}
=
A^\top(K+\lambda I_N)^{-1}\bm y\ .
\label{eq:ridge_gaussian_wstar}
\end{equation}

It is also convenient to introduce the Ridge filter
\begin{equation}
M
=
\lambda(K+\lambda I_N)^{-1}\ .
\label{eq:ridge_filter}
\end{equation}
Using Eq.~\eqref{eq:ridge_gaussian_wstar}, the residual at the minimum is
\begin{align}
\bm y-A\hat{\bm w}
&=
\bm y-K(K+\lambda I_N)^{-1}\bm y
\nonumber\\
&=
\lambda(K+\lambda I_N)^{-1}\bm y
=
M\bm y\ .
\label{eq:ridge_residual_filter}
\end{align}
Similarly,
\begin{equation}
\|\hat{\bm w}\|^2
=
\bm y^\top
(K+\lambda I_N)^{-1}
K
(K+\lambda I_N)^{-1}
\bm y\ .
\end{equation}
Therefore the minimal objective is
\begin{align}
\mathcal E
&=
\frac12\bm y^\top
\left[
M^2
+
\lambda
(K+\lambda I_N)^{-1}
K
(K+\lambda I_N)^{-1}
\right]
\bm y
\nonumber\\
&=
\frac12\bm y^\top
(K+\lambda I_N)^{-2}
\left[
\lambda^2I_N+\lambda K
\right]
\bm y
\nonumber\\
&=
\frac12\bm y^\top
\lambda(K+\lambda I_N)^{-1}
\bm y\ .
\end{align}
Thus
\begin{equation}
{
\mathcal E
=
\frac12\,\bm y^\top M\bm y
}\ .
\label{eq:ridge_quadratic_form}
\end{equation}
This identity is exact at finite $N$ and $P$. It is the Ridge analogue of
the OLS projection formula derived above (see Eq. \eqref{eq:ols_E0_projector}).

For a Gaussian teacher, the conditional distribution of the response is
Gaussian,
\begin{equation}
\bm y\,|\,X
\sim
\mathcal N(0,\Sigma_y),
\qquad
\Sigma_y
=
\sigma^2I_N+vK.
\label{eq:ridge_gaussian_covariance}
\end{equation}
Indeed, conditioning on $X$ (and hence on $A$), and using the independence of $\bm{\beta}$ and $\bm{\epsilon}$ together with $\langle\bm{\beta}\bm{\beta}^\top\rangle=vI_P$ and $\langle\bm{\epsilon}\bm{\epsilon}^\top\rangle=\sigma^2 I_N$, we have
\begin{equation}
\mathrm{Cov}(\bm{y}\mid X)
=\left\langle (A\bm{\beta}+\bm{\epsilon})(A\bm{\beta}+\bm{\epsilon})^\top\right\rangle_{\bm{\beta},\bm{\epsilon}}
=vAA^\top+\sigma^2 I_N
=vK+\sigma^2 I_N\ .
\tag{305}
\end{equation}
Since $A\bm{\beta}$ and $\bm{\epsilon}$ are independent Gaussian vectors, their sum is Gaussian, which gives Eq.~\eqref{eq:ridge_gaussian_covariance}.

Both $\Sigma_y$ and $M$ are functions of $K$ and therefore commute. Using
Eq.~\eqref{eq:ridge_quadratic_form}, the conditional Laplace transform of the
minimal loss is
\begin{align}
\left\langle
\re^{-s\mathcal E}
\right\rangle_{\bm\beta,\bm\epsilon|X}
&=
\left\langle
\exp\left[
-\frac{s}{2}\bm y^\top M\bm y
\right]
\right\rangle_{\bm y|X}
\nonumber\\
&=
\det\left(
I_N+s\Sigma_yM
\right)^{-1/2}.
\label{eq:ridge_conditional_determinant}
\end{align}

This result follows from writing $\bm{y}=\Sigma_y^{1/2}\bm{z}$ with
$\bm{z}\sim\mathcal{N}(0,I_N)$. For a standard Gaussian vector $\bm{z}$ and a
symmetric matrix $B$ such that $I_N+B$ is positive definite, the standard
Gaussian integral gives
\begin{equation}
\left\langle \mathrm{e}^{-\frac{1}{2}\bm{z}^{\top}B\bm{z}}\right\rangle_{\bm{z}}
=
\frac{1}{(2\pi)^{N/2}}
\int_{\mathbb{R}^N}\mathrm{d}\bm{z}\,
\mathrm{e}^{-\frac{1}{2}\bm{z}^{\top}(I_N+B)\bm{z}}
=
\det(I_N+B)^{-1/2}\ .
\tag{306}
\end{equation}
Setting $B=s\Sigma_y^{1/2}M\Sigma_y^{1/2}$ and using
$\det(I+AB)=\det(I+BA)$, we easily obtain \eqref{eq:ridge_conditional_determinant}.

Let $\mu_1,\ldots,\mu_N$ denote the eigenvalues of $K$. Since $K$,
$\Sigma_y$, and $M$ can be simultaneously diagonalized,
Eq.~\eqref{eq:ridge_conditional_determinant} factorizes as
\begin{equation}
\left\langle
\re^{-s\mathcal E}
\right\rangle_{\bm\beta,\bm\epsilon|X}
=
\prod_{i=1}^N
\left[
1+s\,h(\mu_i)
\right]^{-1/2},
\label{eq:ridge_gaussian_product}
\end{equation}
where
\begin{equation}
h(\mu)
=
\frac{
\lambda(\sigma^2+v\mu)
}{
\lambda+\mu
}\ .
\label{eq:ridge_hmu}
\end{equation}

This formula follows from the observation that, since both $\Sigma_y=\sigma^2 I_N+vK$ and $M=\lambda(K+\lambda I_N)^{-1}$ are functions of $K$, they share the same eigenvectors. Thus, if $K\bm{u}_i=\mu_i\bm{u}_i$, then
\begin{equation}
\Sigma_y\bm{u}_i=(\sigma^2+v\mu_i)\bm{u}_i\ ,
\qquad
M\bm{u}_i=\frac{\lambda}{\lambda+\mu_i}\bm{u}_i\ ,
\qquad
\Sigma_y M\bm{u}_i=
\underbrace{\frac{\lambda(\sigma^2+v\mu_i)}{\lambda+\mu_i}}_{h(\mu_i)}
\bm{u}_i \ .
\tag{308}
\end{equation}
Hence, $h(\mu)$ is the eigenvalue of $\Sigma_y M$ associated with an eigendirection of $K$ having eigenvalue $\mu$.

Thus, conditional on the design $X$, the eigenbasis of $K$ also diagonalizes
$\Sigma_y$ and $M$. Writing the response in this basis as
$\bm{y}=U\,\mathrm{diag}\!\left(\sqrt{\sigma^2+v\mu_i}\right)\bm{z}$, with
$z_i\stackrel{\mathrm{i.i.d.}}{\sim}\mathcal{N}(0,1)$, the minimal Ridge
objective becomes
\begin{equation}
E=\frac{1}{2}\bm{y}^{\top}M\bm{y}
=\frac{1}{2}\sum_{i=1}^{N}
\frac{\lambda(\sigma^2+v\mu_i)}{\lambda+\mu_i}\,z_i^2
=\frac{1}{2}\sum_{i=1}^{N} h(\mu_i)\,z_i^2 \ .
\tag{309}
\end{equation}
Since the $z_i$ are independent standard Gaussian variables,
$z_i^2\sim\chi_1^2$ independently. Hence, conditional on $X$, the minimal
Ridge objective is a weighted sum of independent $\chi_1^2$ random variables,
with one contribution from each eigendirection of the Gram matrix and weight
$h(\mu_i)$ determined by the corresponding eigenvalue $\mu_i$.

Taking the logarithm of Eq.~\eqref{eq:ridge_gaussian_product} gives
\begin{equation}
-\frac1P
\log
\left\langle
\re^{-s\mathcal E}
\right\rangle_{\bm\beta,\bm\epsilon|X}
=
\frac{1}{2P}
\sum_{i=1}^N
\log\left[
1+s\,h(\mu_i)
\right]\ .
\label{eq:ridge_linear_statistic}
\end{equation}
For fixed $\lambda>0$ and $s$ in the domain of the Laplace transform,
the function $\log[1+s h(\mu)]$ is bounded on the spectrum. Standard
concentration of linear spectral statistics for Wishart matrices then implies
that fluctuations of Eq.~\eqref{eq:ridge_linear_statistic} are subextensive,
so that its leading $O(P)$ contribution is determined by the limiting
spectral measure of $K$.

Since
\begin{equation}
K=\frac{XX^\top}{P}\ ,
\end{equation}
the relevant Mar\v{c}enko--Pastur aspect ratio is
\begin{equation}
c=\frac{N}{P}=\frac1r\ .
\end{equation}
The limiting spectral measure is
\begin{equation}
\rho_c(\dd\mu)
=
\left(1-\frac1c\right)_+\delta_0(\dd\mu)
+
\frac{
\sqrt{(\mu_+-\mu)(\mu-\mu_-)}
}{
2\pi c\,\mu
}
\dd\mu,
\qquad
\mu_\pm=(1\pm\sqrt c)^2\ .
\label{eq:MP_ridge_check}
\end{equation}
We therefore obtain from the continuum limit of \eqref{eq:ridge_linear_statistic} the expression
\begin{equation}
{
\Phi_{\rm Ridge}(s)
=
\frac{1}{2r}
\int
\rho_{1/r}(\dd\mu)\,
\log
\left[
1+s
\frac{
\lambda(\sigma^2+v\mu)
}{
\lambda+\mu
}
\right]
}\ .
\label{eq:ridge_replica_free_phi}
\end{equation}
For a Gaussian teacher, Eq.~\eqref{eq:ridge_replica_free_phi} provides an
independent one-dimensional representation of the same scaled cumulant
generating function obtained from the replica saddle-point equations.

\paragraph{OLS limit.}
As a first check, consider $\lambda\to0^+$ with $r<1$. For every
$\mu>0$, $h(\mu)\to0$, whereas
\begin{equation}
h(0)=\sigma^2\ .
\end{equation}
The Mar\v{c}enko--Pastur measure has an atom of mass $1-r$ at the origin,
and therefore
\begin{equation}
\Phi_{\rm Ridge}(s)
\underset{\lambda\to0^+}{\longrightarrow}
\frac{1-r}{2r}
\log(1+s\sigma^2)\ .
\label{eq:ridge_check_OLS}
\end{equation}
For $\sigma^2=1$, this is precisely the OLS result derived above (see Eq. \eqref{eq:ols_replica_phi_final})
\begin{equation}
\Phi_0(s)
=
\frac{1-r}{2r}\log(1+s)\ .
\end{equation}

\paragraph{Identification of the Ridge order parameters.}
A more detailed check follows from the resolvent of the
Mar\v{c}enko--Pastur spectrum. Define
\begin{equation}
G_{\rm MP}(\tau)
=
\int
\frac{
\rho_{1/r}(\dd\mu)
}{
\mu+\tau
}\ .
\label{eq:ridge_MP_resolvent}
\end{equation}
For the Mar\v{c}enko--Pastur law it satisfies
\begin{equation}
\tau G_{\rm MP}^2
+
(r\tau+r-1)G_{\rm MP}
-r
=
0\ .
\label{eq:ridge_MP_resolvent_quadratic}
\end{equation}
On the replica side, the Ridge saddle-point equations imply
\begin{equation}
\hat\chi
=
\frac{1}{2r(1+\chi)},
\qquad
\chi
=
\frac{1}{\lambda+2\hat\chi}\ .
\label{eq:ridge_check_sp}
\end{equation}
Eliminating $\hat\chi$ gives
\begin{equation}
\lambda r\chi^2
+
(\lambda r+1-r)\chi
-r
=
0\ .
\label{eq:ridge_check_chi_quadratic}
\end{equation}
Equations~\eqref{eq:ridge_MP_resolvent_quadratic} and
\eqref{eq:ridge_check_chi_quadratic} are equivalent under
\begin{equation}
{
\lambda G_{\rm MP}(\lambda)
=
\frac{1}{1+\chi},
\qquad
\hat\chi
=
\frac{\lambda G_{\rm MP}(\lambda)}{2r}
}\ .
\label{eq:ridge_resolvent_identification}
\end{equation}
Thus the Ridge order parameters $\chi$ and $\hat\chi$ are directly related
to the resolvent of the design spectrum. Indeed, using
\begin{equation}
G_{\rm MP}(\lambda)=\frac{1}{\lambda(1+\chi)}
\end{equation}
in Eq.~\eqref{eq:ridge_MP_resolvent_quadratic} gives
\begin{align}
0
&=\lambda G_{\rm MP}^2+(r\lambda+r-1)G_{\rm MP}-r \nonumber\\
&=\frac{1}{\lambda(1+\chi)^2}
 +\frac{r\lambda+r-1}{\lambda(1+\chi)}-r \ .
\end{align}
Multiplying by $\lambda(1+\chi)^2$ and rearranging, one obtains
\begin{equation}
r\lambda\chi^2+(r\lambda+1-r)\chi-r=0\ ,
\end{equation}
which is precisely Eq.~\eqref{eq:ridge_check_chi_quadratic}. Moreover, the first saddle-point equation immediately gives
\begin{equation}
\widehat{\chi}
=\frac{1}{2r(1+\chi)}
=\frac{\lambda G_{\rm MP}(\lambda)}{2r}\ .
\end{equation}
Thus, the replica saddle-point equations encode the same spectral self-consistency as the Mar\v{c}enko--Pastur resolvent: the susceptibility $\chi$ is related directly to the spectral response of the random design, while its conjugate parameter $\widehat{\chi}$ is fixed by the same resolvent.

\paragraph{Mean minimal loss.}
Differentiating Eq.~\eqref{eq:ridge_replica_free_phi} at $s=0$ gives
\begin{equation}
\lim_{P\to\infty}
\frac{\langle\mathcal E\rangle}{P}
=
\Phi_{\rm Ridge}'(0)
=
\frac{1}{2r}
\int\rho_{1/r}(\dd\mu)\,h(\mu)\ .
\end{equation}
For $\sigma^2=1$,
\begin{equation}
h(\mu)
=
\lambda v
+
\frac{
\lambda(1-\lambda v)
}{
\lambda+\mu
}\ ,
\end{equation}
and hence
\begin{equation}
\Phi_{\rm Ridge}'(0)
=
\frac{1}{2r}
\left[
\lambda v
+
\lambda(1-\lambda v)
G_{\rm MP}(\lambda)
\right]\ .\label{eqPhiprimeRidgewithG}
\end{equation}
Using now Eq.~\eqref{eq:ridge_resolvent_identification} and $\widehat{\chi}=1/[2r(1+\chi)]$\ , Eq.~\eqref{eqPhiprimeRidgewithG} can be rewritten as
\begin{align}
\Phi'_{\rm Ridge}(0)
&=\frac{\lambda v}{2r}+(1-\lambda v)\widehat{\chi}
=\widehat{\chi}+\lambda v\left(\frac{1}{2r}-\widehat{\chi}\right) \nonumber\\
&=\widehat{\chi}+\lambda v\,\chi\widehat{\chi}\ ,
\end{align}
where in the last equality we used
$1/(2r)-\widehat{\chi}=\chi\widehat{\chi}$. This is equivalent to the average Ridge loss obtained from the replica
calculation (see Eq. \eqref{FINALeqAverageRSS} upon using
$1-\lambda\chi=2\chi\widehat{\chi}$).

We stress that the Gaussian assumption is essential to the determinant reduction above.
For a general teacher prior, $\bm y|X$ is no longer Gaussian and the average
in Eq.~\eqref{eq:ridge_conditional_determinant} cannot in general be reduced
to a determinant. The replica formulation is therefore still required for
the arbitrary priors considered in the main calculation.

\end{document}